\documentstyle[12pt]{article}
\begin{document}
\pagestyle{empty}
\hspace*{12.4cm}IU-MSTP/30 \\
\hspace*{13cm}hep-th/9810189 \\
\hspace*{13cm}October, 1998
\begin{center}
 {\Large\bf Quantum Exchange Algebra and Exact Operator Solution 
of $A_2$-Toda Field Theory}
\end{center}

\vspace*{1cm}
\def\thefootnote{\fnsymbol{footnote}}
\begin{center}{\sc Y. Takimoto,} 
{\sc H. Igarashi,} {\sc H. Kurokawa} 
and {\sc T. Fujiwara}$^1$ 
\end{center}
\vspace*{0.2cm}
\begin{center}
{\em Graduate School of Science and Engineering,
Ibaraki University, Mito 310-8512, Japan}\\
{\em $\ ^{1}$ Department of Mathematical Sciences, Ibaraki University,
Mito 310-8512, Japan}\\
\end{center}
\vfill
\begin{center}
{\large\sc Abstract}
\end{center}
\noindent
Locality is analyzed for Toda field theories by noting novel chiral 
description in the conventional nonchiral formalism. It is shown that 
the canonicity of 
the interacting to free field mapping described by the classical 
solution is automatically guaranteed by the locality. Quantum Toda 
theories are investigated by applying the method of free field 
quantization. We give Toda exponential operators associated with 
fundamental weight vectors as bilinear forms of chiral fields 
satisfying characteristic quantum exchange algebra.  It is shown 
that the locality leads to nontrivial relations among the 
${\cal R}$-matrix and the expansion coefficients of the 
exponential operators. The Toda exponentials are obtained for 
$A_2$-system by extending the algebraic method developed for 
Liouville theory. The canonical commutation relations and the 
operatorial field equations are also examined.



\newpage
\pagestyle{plain}

\section{Introduction}
\label{sec:intro}

Toda field theories in two dimensions are described 
by the characteristic exponential type interactions. 
One can not naively apply ordinary quantization 
procedure for them since there is no vacuum for finite 
field configurations and the couplings between the 
fluctuations around any classical background tend to 
zero as it approaches to the potential minima. In spite 
of such a seemingly complicated feature, they are 
completely integrable and the exact solution at the 
classical level is known \cite{ls}. It can be 
expressed in terms of free fields which are 
related to Toda fields by canonical 
transformation. These free fields can be regarded as 
normal modes of the interacting fields. Thus the 
quantum theory of Toda fields can be defined by 
imposing canonical commutation relations on the 
free fields. 

The simplest Toda theory associated with the Lie 
algebra $A_1$ is Liouville theory, for which extensive 
canonical approaches have been developed so far \cite{gn,bcgt}. 
There are two different operator approaches 
for quantum Liouville theory. The one advocated by 
Gervais and Neveu \cite{gn} is based on free field 
realization of stress tensor. 
Basic building blocks are chiral vertex operators 
or Bloch wave solutions, which satisfy characteristic 
exchange algebra \cite{gn84} related with $U_q(sl_2)$ quantum 
group symmetry \cite{Babe,cgr}. Liouville exponential operators 
can be constructed from them correspondingly to 
half-integral spin representation of $U_q(sl_2)$. 
We refer to this approach as chiral scheme since left- and 
right-moving variables are taken as independent. 

The other approach 
was proposed by Braaten, Curtright and Thorn \cite{bcgt}, 
who noted that the B\"acklund transformation 
relating Liouville field with free one is a canonical 
mapping and it can be used to define quantum theory as 
mentioned above. They gave exact expressions of some 
basic operators of the theory. In this approach left- 
and right-moving variables are not independent due to 
symmetrical assignment of zero-mode variables of the 
canonical free fields. We refer to this as vector scheme. 

Construction of Liouville exponential operators of 
arbitrary conformal weights were investigated by Otto and 
Weigt \cite{ow} basically along the line of thought of 
ref. \cite{bcgt} 
but with a different parametrization of the 
classical solution \cite{dhj}. Assuming most general expansion 
of Liouville exponential operators as power series of 
screening charge consistently with conformal invariance, 
they showed that the Liouville exponential operators are 
determined almost uniquely by imposing locality condition. 
They carried out the analysis to third order in cosmological 
constant and inferred exact operator solution, which can 
be interpreted as a quantum deformation of 
the classical expressions. Gervais and Schnittger \cite{gs93,gs94} 
arrived at exact solution by extending the formalism of 
chiral vertex operators \cite{cgr} to continuous spin representation. 
They made detailed comparision between the chiral and 
vector schemes and established the equivalence of their 
operator solution with the conjecture given in ref. 
\cite{ow}.\footnote{We thank J. Schnittger for useful 
comments on this point.} Three of the present authors 
also carried out the analysis within the vector scheme 
by applying the algebraic method developed in ref. \cite{gs93} 
and reconfirmed the operator solution to all order 
of cosmological constant \cite{fit96}. 

It is very natural to expect that similar development 
can also be achieved for quantum Toda field theories. 
Such is of special interest since they incorporate 
$W$-symmetry \cite{wsymm} and serve as gravity sector of 
extended conformal field theories. In fact 
Toda systems have been investigated extensively with 
emphasis on their extended conformal structures 
\cite{Babe,bg,bg89,bfo,wgeom}. 
In particular Bilal and Gervais \cite{bg,bg89} have shown that 
the apporach of chiral vertex operators developed for 
Liouville theory can be extended to Toda theories. 
Construction of Toda exponential operators of arbitrary 
conformal weights as in Liouville theory, however, has 
not been achieved for general Toda 
theories.\footnote{Fedoseev and Leznov \cite{fl} have 
investigated generalized Toda system within vector scheme 
and have consturcted exponential operators of particular 
types.}

Three of the present authors have recently extended the 
approach of ref. \cite{ow} to $A_2$-Toda system and obtained 
arbitrary Toda exponential operators to fourth order in 
the cosmological constant by analyzing directly the 
locality conditons within vector scheme, giving a 
conjecture for exact operator solution of $A_2$-Toda 
theory \cite{fit98}. The method is rather restricted to $A_2$ case 
and becomes intractable as one goes to higher orders 
though the basic ideas undelying the arguments presented 
there could equally well be applied for general Toda 
systems. The purpose of this paper is to investigate 
it from a more general setting that is not specialized 
to a particular Toda system as possible as we can, and to 
generalize the algebaric method 
developed for Liouville theory \cite{gs93,fit96} to higher rank 
systems containing more than one screening charges. 
Since chiral approach is more convenient 
than vectorial one, we shall show that a chiral structure 
can be defined for Toda systems written in vector scheme 
without introducing any extra degrees of freedom, and 
mostly work in the chiral schematic description. This not 
only explains the equivalence between the two approaches 
in a direct manner but also enables us to restate the 
locality of Toda exponential operators as a property of 
the ${\cal R}$-matrix \cite{gn84,gs93}. The canonical commutation 
relations will turn out to be a straightforward consequence 
of the locality. 

As stressed in ref. \cite{fit98}, it is very convenient to consider 
Toda exponential operators associated with the fundamental 
weights. They contain screening charges and a free 
field vertex operator which are mutually commuting, hence 
no ordering problem arise in the operator products. A full 
set of those exponential operators is sufficient to 
reconstruct not only arbitrary  exponential operators
but also the local Toda field operator itself.
The generalization of the algebraic method of ref. \cite{fit96} to 
$A_2$-system is not straightforward due to the fact that 
the exchange algebra of the screening charges can not be 
realized by a simple quantum mechanical system as in 
Liouville theory. We need to truncate the full screening 
charges to solve the locality constraints so that the 
algebra of the truncated operators allows a quantum 
mechanical realization. It will be shown that this can be 
done and the Toda exponential operators can be determined 
up to a constant related to the arbitrariness of the 
cosmological constant. General exponential operators such 
as the Toda potential can be obtained as operator products 
from these operators. That the local Toda field satisfies 
the operatorial field equation can be established 
in a straighforward way. Strictly speaking, the 
locality should be reexamined with 
the operator solution with the full screening charges 
since one can not a priori expect locality to be built-in. 
It turns out to be a rather hard problem to check this 
directly with the full operator expression. One way to 
achieve this is to establish the property of the 
${\cal R}$-matrix from which the locality follows. This 
can be done explicitly for Liouville theory and partially 
for $A_2$-Toda case. We think it to be a technical aspect 
of the theory and will be solved positively. 

This paper is organized as follows: In sect. 2 we argue 
canonical structure of classical Toda theories. Chiral 
schematic description of vectorial theories is introduced. 
Canonicity of the mapping between Toda system and the free 
theory defined by the classical solution is formulated 
as constraints for the $r$-matirx, which are examined 
explicitly for $A_1$ and $A_2$. Free field quantization of 
$A_2$-Toda field is described in sect. 3. It is shown that
locality leads to nontrivial relations between the 
${\cal R}$-matrix and the coefficients appearing in the 
Toda exponential operator. That the locality automatically 
guarantees the canonical commutation relations is also 
argued. In sect. 4 the exponential operators of 
$A_2$-system are obtained by extending the algebraic 
approach of ref. \cite{fit96}. Sect. 5 deals with the operatorial 
field equation. Closed forms of the ${\cal R}$-matrix 
and their property discussed in sect. 3 are established 
for some special cases in sect. 6. The final section is 
devoted to discussions. In Appendices A and B we summarize 
Poisson brackets among the screening charges, quadratic 
Poisson algebra satisfied by the chiral fields of 
$A_2$-system and the basic quantum exchange algebra. 
Appendix C provides an elementary proof for a function 
introduced in sect. 6. 

\section{Canonical structure of classical Toda field theory}
\label{sec:classical toda}

Though we will be mostly concerned with $A_2$-Toda field theory, we 
begin with some discussions on the exact solution of classical Toda 
theories associated with an arbitrary simple Lie algebra ${\cal G}$
of rank $r$. It is described by the action
\begin{eqnarray}
  \label{cl-act}
  S=\frac{1}{\gamma^2}\int_{-\infty}^{+\infty}d\tau
  \int_0^{2\pi}d\sigma \Biggl(\frac{1}{2}\partial_\alpha\varphi\cdot
  \partial^\alpha\varphi-\mu^2\sum_{a=1}^r
  {\rm e}^{\alpha^a\cdot\varphi}\Biggr) ~,
\end{eqnarray}
where $\varphi$ is the $r$-component Toda field and $\alpha^a$ 
($a=1,\cdots,r$) stand for the simple roots of ${\cal G}$. The coupling 
constant $\gamma$ is a free parameter, which is determined by the 
requirement of conformal invariance in the presence of conformal 
matters. 

The Toda field equations 
\begin{eqnarray}
  \label{teq}
  \partial_\alpha\partial^\alpha\varphi-\mu^2\sum_{a=1}^r \alpha^a
  {\rm e}^{\alpha^a\cdot \varphi}=0
\end{eqnarray}
can be solved exactly by Lie algebraic method \cite{ls,Babe,btb}. We 
work in the Cartan-Weyl 
basis and denote the generators of the Cartan subalgebra of ${\cal G}$ by 
$H_k$ ($k=1,\cdots,r$) and the step operators corresponding to the simple 
root $\alpha^a$ by $E_{\pm\alpha^a}$. They satisfy $[H,E_{\pm\alpha^a}]=
\pm\alpha^aE_{\pm\alpha^a}$, $[E_{\alpha^a},E_{-\alpha^b}]=\delta^{ab}\alpha^a
\cdot H$, where $H$ is an $r$-component vector with $H_k$ as the $k$-th 
component. Let $\lambda^a$ ($a=1,\cdots,r$) be the fundamental weights 
satisfying $2\lambda^a\cdot\alpha^b/(\alpha^b)^2=\delta^{ab}$ and 
$|\lambda^a\rangle$ be the corresponding highest weight vector characterized 
by $H|\lambda^a\rangle=\lambda^a|\lambda^a\rangle$, 
$E_{\alpha^b}|\lambda^a\rangle=0$. We also introduce the adjoint vectors 
$\langle\lambda^a|$ satisfying $\langle\lambda^a|\lambda^b\rangle
=\delta^{ab}$, $\langle\lambda^a|H=\alpha^a\langle\lambda^a|$ and 
$\langle\lambda^a|E_{-\alpha^b}=0$. Then the exact solution to (\ref{teq}) 
is given by
\begin{eqnarray}
  \label{exacsol}
   {\rm e}^{\lambda^a\cdot\varphi(x)}=\frac{{\rm e}^{\lambda^a\cdot\psi(x)}}{%
     \langle\lambda^a|M_+(x^+)M_-^{-1}(x^-)|\lambda^a\rangle}~,
\end{eqnarray}
where $\psi(x)=\psi_+(x^+)+\psi_-(x^-)$ with $\psi_\pm(x^\pm)$ being 
arbitrary functions of the light-cone variables $x^\pm=\tau\pm\sigma$. 
It is identified with the canonical 
free field. $M_\pm(x^\pm)$ are defined by the 
differential equations
\begin{eqnarray}
  \label{meq}
  \partial_\pm M_\pm(x^\pm)=\mp\frac{\mu}{2}\sum_{a=1}^r 
  V_a^\pm(x^\pm)E_{\pm\alpha^a}M_\pm(x^\pm)~,
\end{eqnarray}
where $V^\pm_a(x^\pm)={\rm e}^{\alpha^a\cdot\psi_\pm(x^\pm)}$ are the 
classical vetices. 

To ensure the periodic boundary condition on $\varphi$, we assume the 
same periodicity for $\psi$. The left- and the right-moving modes 
$\psi_\pm$, however, are not $2\pi$ periodic. To see this, we note the 
normal mode expansions 
\begin{eqnarray}
  \label{nme}
  \psi_\pm(x^\pm)=\frac{\gamma}{2}Q+\frac{\gamma}{4\pi}Px^\pm
  +\frac{i\gamma}{\sqrt{4\pi}}\sum_{n\neq 0}\frac{1}{n}a_n^{(\pm)}
  {\rm e}^{-inx^\pm} ~.
\end{eqnarray}
Then $\psi_\pm$ satisfy the periodicity $\psi(x^\pm\pm2\pi)
=\psi_\pm(x^\pm)\pm\displaystyle{\frac{\gamma}{2}P}$. To establish the 
$2\pi$ periodicity of $\varphi$, we need to show the invariance 
of the denominator of the rhs of (\ref{exacsol}). This can be seen from 
the periodicity of $M_\pm$
\begin{eqnarray}
  \label{mperiod}
  M_\pm(x^\pm\pm2\pi)={\rm e}^{\frac{\gamma}{2}P\cdot H}M_\pm(x^\pm)
  {\rm e}^{-\frac{\gamma}{2}P\cdot H}~,
\end{eqnarray}
which can be deduced from (\ref{meq}). 

We can integrate (\ref{meq}) iteratively by noting the periodicity 
(\ref{mperiod}) as 
\begin{eqnarray}
  \label{mpm}
  M_+(x^+)&=&\sum_{n=0}^\infty\Biggl(-\frac{\mu}{2}\Biggr)^n
  \sum_{a_1,\cdots,a_n}A_{a_1\cdots a_n}(x^+)
  E_{\alpha^{a_1}}\cdots E_{\alpha^{a_n}}~, \\
  M_-^{-1}(x^-)&=&\sum_{n=0}^\infty\Biggl(-\frac{\mu}{2}\Biggr)^n
  \sum_{a_1,\cdots,a_n}B_{a_1\cdots a_n}(x^-)
  E_{-\alpha^{a_n}}\cdots E_{-\alpha^{a_1}}~,
\end{eqnarray}
where the screening charge $A_{a_1\cdots a_n}$ is defined 
\begin{eqnarray}
  \label{a}
  A_{a_1\cdots a_n}(x)\hskip -.2cm&=&\hskip -.2cm 
  C_{\alpha^{a_1}+\cdots+\alpha^{a_n}}
  C_{\alpha^{a_2}+\cdots+\alpha^{a_n}}\cdots C_{\alpha^{a_n}}
  \int_0^{2\pi}\hskip -.15cm dy_1
  {\cal E}_{\alpha^{a_1}+\cdots+\alpha^{a_n}}(x-y_1) 
  V^+_{a_1}(y_1) \nonumber\\  
  \hskip -.2cm&&\hskip -.2cm \times \int_0^{2\pi}\hskip -.15cm dy_2
  {\cal E}_{\alpha^{a_2}+\cdots+\alpha^{a_n}}(y_1-y_2)V^+_{a_2}(y_2)
  \cdots\int_0^{2\pi\hskip -.15cm }dy_n{\cal E}_{\alpha^{a_n}}(y_{n-1}-y_n)
  V^+_{a_n}(y_n)~. 
\end{eqnarray}
In (\ref{a}) we have introduced for an arbitrary vector $\beta$ in the 
root space
\begin{eqnarray}
  \label{ce}
  C_\beta=\Biggl(2{\rm sinh}\frac{\gamma}{4}\beta\cdot P\Biggr)^{-1}~, \qquad
  {\cal E}_\beta(x)=\exp \frac{\gamma}{4} \beta\cdot P \epsilon(x)
\end{eqnarray}
with $\epsilon(x)$ being stair-step function defined by $\epsilon(x)=
1$ for $0<x<2\pi$ and $\epsilon(x+2\pi)=\epsilon(x)+2$. The screening 
charges satisfy the quasi-periodicity
\begin{eqnarray}
  \label{quasip}
  A_{a_1\cdots a_n}(x^++2\pi)={\rm e}^{\frac{\gamma}{2}(\alpha^{a_1}
    +\cdots+\alpha^{a_n})\cdot P}A_{a_1\cdots a_n}(x^+) ~.
\end{eqnarray}
The screening charges in the right-moving sector $B_{a_1\cdots a_n}$ 
are obtained by replacing $V_a^+$ with $V_a^-$ in the rhs of (\ref{a}). One 
can verify from (\ref{a}) and (\ref{ce}) that $A_{a_1\cdots a_n}$ satisfy 
the differential equations
\begin{eqnarray}
  \label{aseq}
  \partial_+A_a(x^+)=V_a^+(x^+)~, \qquad 
  \partial_+A_{a_1\cdots a_n}(x^+)=V_{a_1}^+(x^+)A_{a_2\cdots a_n}(x^+)~.
\end{eqnarray}
That $M_+$ satisfies (\ref{meq}) can be seen easily from (\ref{mpm}) and 
(\ref{aseq}). Similar argument also applies to $B_{a_1\cdots a_n}$ and 
$M_-$. 

Putting (\ref{mpm}) into the rhs of (\ref{exacsol}), we can express 
the classical solution in the following form
\begin{eqnarray}
  \label{exacsol2}
  {\rm e}^{\lambda^a\cdot\varphi(x)}=\frac{{\rm e}^{\lambda^a\cdot\psi(x)}}{%
    1+\displaystyle{\sum_{n=1}^{\infty}\Biggl(\frac{\mu^2}{4}\Biggr)^n
    \sum_{\{a\}_n,\{b\}_n}C^a_{\{a\}_n;\{b\}_n}
    A_{\{a\}_n}(x^+)B_{\{b\}_n}(x^-)}}~,
\end{eqnarray}
where $\{a\}_n$ stands for the set of $n$ ordered indices $a_1\cdots a_n$ and 
$C^a_{\{a\}_n;\{b\}_n}$ are numerical constants given by
\begin{eqnarray}
  \label{C}
  C^a_{\{a\}_n;\{b\}_n}=\langle\lambda^a|E_{\alpha^{a_1}}\cdots 
  E_{\alpha^{a_n}}
  E_{-\alpha^{b_n}}\cdots E_{-\alpha^{b_1}}|\lambda^a\rangle ~.
\end{eqnarray}
These coefficients are determined only from the Lie algebra ${\cal G}$ 
and vanish for sufficiently large $n$ due to the finite dimensionality 
of the fundamental representations. Furthermore, they are symmetric, i.e.,  
$C^a_{\{a\}_n;\{b\}_n}=C^a_{\{b\}_n;\{a\}_n}$, and satisfy 
\begin{eqnarray}
  \label{C0}
  C^a_{\{a\}_n;\{b\}_n}=0 \quad {\rm unless}\quad 
  \alpha^{a_1}+\cdots+\alpha^{a_n}=\alpha^{b_1}+\cdots+\alpha^{b_n}~.
\end{eqnarray}
We thus obtain the generalization of the well-known classical solution to 
Liouville equation which corresponds to $A_1$ case. For $A_2$-Toda theory the 
solution is explicitly given by 
\begin{eqnarray}
  \label{a2csol}
  {\rm e}^{\lambda^a\cdot\varphi(x)}=\frac{{\rm e}^{\lambda^a\cdot\psi(x)}}{%
    1+\displaystyle{\frac{\mu^2}{4}A_a(x^+)B_a(x^-)
    +\biggl(\frac{\mu^2}{4}\biggr)^2A_{a\bar a}(x^+)B_{a\bar a}(x^-)}} ~, 
  \qquad
  (a=1,2)
\end{eqnarray}
where the convention $\bar 1=2$, $\bar 2=1$ is employed for the indices, 
specifically for $A_2$ case.

The remarkable property of Toda theories is that (\ref{exacsol2}) defines a 
canonical transformation from $\psi$ to $\varphi$. In other words the 
fundamental Poisson brackets among the canonical variables $\varphi$ and 
$\pi_\varphi\equiv\displaystyle{\frac{1}{\gamma^2}\partial_\tau\varphi}$
are guaranteed by the canonical pairs of the free fields  $\psi$ and 
$\pi_\psi\equiv\displaystyle{\frac{1}{\gamma^2}\partial_\tau\psi}$, or 
equivalently by the Poisson brackets
\begin{eqnarray}
  \label{poisbra}
  \{\psi_k(x),\psi_l(x')\}=-\frac{\gamma^2}{4}(\epsilon(x^+-x'{}^+)
  +\epsilon(x^--x'{}^-))\delta_{kl} ~. \qquad
  (k,l=1,\cdots,r)
\end{eqnarray}
One can be convinced of this by noting that the conserved higher-spin 
currents generating the extended conformal symmetry can be written in 
the same form by the substitution $\partial_\pm\varphi\rightarrow
\partial_\pm\psi$ \cite{Babe,bg}. This can be shown from the general ground 
of Toda theories. The simplest is the stress tensor, for which one 
can verify this most directly as 
\begin{eqnarray}
  \label{st}
  T_{\pm\pm}&=&\frac{1}{\gamma^2}
  (\partial_\pm\varphi\cdot\partial_\pm\varphi
  -2\rho\cdot\partial_\pm^2\varphi) \nonumber\\
  &=&\frac{1}{\gamma^2}
  (\partial_\pm\psi\cdot\partial_\pm\psi
  -2\rho\cdot\partial_\pm^2\psi) ~,
\end{eqnarray}
where $\rho$ is assumed to satisfy $\alpha^a\cdot\rho=1$ for any simple 
root and is explicitly given by
\begin{eqnarray}
  \label{rho}
  \rho=\sum_a\frac{2\lambda^a}{(\alpha^a)^2} ~.
\end{eqnarray}
Similar situation occurs for higher-spin conserved currents. 
In terms of canonical variables these two expressions, the one in
terms of interacting Toda fields and the other in terms of the canonical 
free fields, coincide each other except for the terms containing the 
cosmological constant $\mu^2$. They form a closed algebra in the 
sense of Poisson bracket, which is insensitive to the parameter $\mu^2$. 
This almost implies the free field Poisson bracket
\begin{eqnarray}
  \label{dpoisbra}
  \{\partial_\pm\psi_{k\pm}(x^\pm),\partial_\pm\psi_{l\pm}(x'{}^\pm)\}
  =\frac{\gamma^2}{2}\partial_\pm\delta(x^\pm
  -x'{}^\pm)\delta_{kl}~. \qquad (k,l=1,\cdots,r)
\end{eqnarray}
Later in this section we will show more rigorously the canonicity 
between the interacting and the free fields for the $A_1$ and $A_2$ cases. 

Eq. (\ref{dpoisbra}) does not contain the zero-modes of $\psi_\pm$. 
To fix the zero-mode dependence we require that 
the conformal dimension of ${\rm e}^{\lambda^a\cdot\varphi}$ coincides with 
that of ${\rm e}^{\lambda^a\cdot\psi}$. By treating the left- and the 
right-moving modes symmetrically we arrive at (\ref{poisbra}) with the 
normal mode expansions (\ref{nme}). We refer to the left-right symmetrical 
description as vector scheme as mentioned in the introduction. 

One may consider the left- and the right-moving modes as independent. 
Hence the zero-modes $P_\pm$ of $\partial_\pm\psi_\pm$ are independent 
dynamical variables. The zero-modes of $\psi_\pm$ are determined from the 
requirement that ${\rm e}^{\lambda^a\cdot\psi_+}$ 
and ${\rm e}^{\lambda^a\cdot\psi_-}$, respectively, are of conformal 
weight $(\Delta,0)$ and $(0,\Delta)$, where $2\Delta$ is the conformal 
scaling dimension of ${\rm e}^{\lambda^a\cdot\varphi}$. Then the 
zero-modes $Q_\pm$ of $\psi_\pm$ which are canonical conjugate to 
$P_\pm$ are also independent.
We call such treatment as chiral scheme. The normal mode expansions in 
the chiral scheme are obtained from (\ref{nme}) by the replacement 
$Q,~P\rightarrow2Q_\pm,~P_\pm$. Since the zero-mode variables 
are doubled in the chiral scheme, the phase space is somewhat enlarged 
than that of the vector scheme. The restriction of the phase 
space of the chiral scheme to that of the vector scheme can be consistently 
carried out by imposing the conditions $Q=Q_++Q_-$ and $P=P_+=P_-$. This 
can be understood by noting that only the combination $Q_++Q_-$ appears 
in (\ref{exacsol2}) and the periodic boundary condition on $\varphi$ 
implies $P_+-P_-=0$. 

Though the chiral scheme involves extra variables, it 
has an advantageous point that the left- and the right-moving variables 
appearing in (\ref{exacsol2}) such as 
${\rm e}^{\lambda^a\cdot\psi_\pm(x^\pm)}$, $A_{a_1\cdots a_n}(x^+)$, 
$B_{a_1\cdots a_n}(x^-)$ form closed algebras under Poisson brackets. 
This greatly facilitates the canonical analysis. In the vector scheme, 
however, the two chiral sectors are not completely independent since 
they have the zero-modes in common, and the chiral components do not 
form closed algebra. At the quantum level these chiral fields 
satisfy characteristic exchange algebra \cite{gn84} and quantum group structure
arises naturally in the chiral scheme \cite{Babe}, whereas these are not 
manifest in the vector scheme. 

One might think that the useful chiral structure were lost at the 
expense of dealing only with the physical variables. But this is not 
the case. We can fit the chiral structure in the vector scheme by 
using a trick of rearranging the zero-mode. Let us redefine 
$\psi_\pm(x^\pm)$ given by (\ref{nme}) as 
\begin{eqnarray}
  \label{cnme}
    \psi_\pm(x^\pm)=\gamma Q+\frac{\gamma}{4\pi}Px^\pm
  +\frac{i\gamma}{\sqrt{4\pi}}\sum_{n\neq 0}\frac{1}{n}a_n^{(\pm)}
  {\rm e}^{-inx^\pm} ~.
\end{eqnarray}
They satisfy the Poisson brackets
\begin{eqnarray}
  \label{cpoisbr}
  \{\psi_{k\pm}(x^\pm),\psi_{l\pm}(x'{}^\pm)\}=-\frac{\gamma^2}{4}
  \epsilon(x^\pm-x'{}^\pm)\delta_{kl}~.
\end{eqnarray}
We also redefine $V_a^\pm$, $A_{\{a\}_n}$ and $B_{\{a\}_n}$ by using 
(\ref{cnme}) instead of (\ref{nme}) for $\psi_\pm$. 
Since only the products of the left- and the right-moving variables with the 
equal exponential dependence on $Q$ appear in the rhs of (\ref{exacsol2}), 
we may define a product denoted by $\star$. 
It is the weighted multiplication of an 
arbitrary left-moving variable $L$ and an arbitrary right-moving one $R$,
\begin{eqnarray}
  \label{starp}
  L\star R\equiv L{\rm e}^{-\gamma\omega\cdot Q}R~, 
\end{eqnarray}
where $\omega$ is chosen so as for the Poisson bracket between 
$L{\rm e}^{-\gamma\omega\cdot Q}$
and ${\rm e}^{-\gamma\omega\cdot Q}R$ to vanish. This satisfies the 
product rule
\begin{eqnarray}
  \label{prodrule}
  L\star R L'\star R'=LL'\star RR'~.
\end{eqnarray}
In terms of the $\star$-product an arbitrary free field exponential can be 
written as 
\begin{eqnarray}
  \label{fexp}
  {\rm e}^{\beta\cdot\psi(x)}={\rm e}^{\beta\cdot\psi_+(x^+)}\star 
  {\rm e}^{\beta\cdot\psi_-(x^-)}~.
\end{eqnarray}
A remarkable property of the $\star$-product is that for any pair 
$L\star R$ and $L'\star R'$, their Poisson bracket satisfies 
\begin{eqnarray}
  \label{pbialg}
  \{L\star R,L'\star R'\}=\{L,L'\}\star RR'+LL'\star\{R,R'\} ~,
\end{eqnarray}
Even if the left- and the right-moving variables develop nonvanishing 
Poisson brackets, this implies that they can be regarded as 
independent variables under the $\star$-product. By considering the 
Poisson bracket between (\ref{fexp}) and $L\star R$, and then computing 
the derivative with respect to $\beta$ at $\beta=0$, we obtain 
\begin{eqnarray}
  \label{psipb}
  \{\psi(x),L\star R\}=\{\psi_+(x^+),L\}\star R+L\star\{\psi_-(x^-),R\}~.
\end{eqnarray}
Hence $\psi(x)$ can be regarded as $\psi_+(x^+)\star1+1\star\psi_-(x^-)$ 
in the Poisson bracket. The independence of the left- and the right-moving 
variables in the chiral scheme corresponds to $\{\psi_+(x^+)\star1,1\star
\psi_-(x^-)\}=0$. Similarly, the stress tensors $T_{++}$ and 
$T_{--}$, for instance, can be regarded as $T_{++}\star1$ and 
$1\star T_{--}$, respectively. As we will see in the next section, the 
chiral structure in the vector scheme can be extended to quantum theory. 

We now turn to the analysis of the intimate connection of quadratic 
Poisson algebra satisfied by the chiral fields with the canonical 
structure of the Toda theory.\footnote{This has been investigated in 
refs. \cite{Babe,btb} within chiral scheme for general Toda theories.} 
Using the redefinition (\ref{cnme}) for 
$\psi_\pm$, we put (\ref{exacsol2}) into the following form
\begin{eqnarray}
  \label{cexacsol2}
  {\rm e}^{-\lambda^a\cdot\varphi(x)}
  =\sum_{n=0}^{\infty}\Biggl(\frac{\mu^2}{4}\Biggr)^n
  \sum_{\{a\}_n,\{b\}_n}C^a_{\{a\}_n;\{b\}_n}
  \psi^+_{\{a\}_n}(x^+)\star \psi^-_{\{b\}_n}(x^-)~,
\end{eqnarray}
where the chiral fields $\psi^\pm_{\{a\}_n}$ are defined by
\begin{eqnarray}
  \label{psipm}
  \psi^+_{\{a\}_n}(x^+)={\rm e}^{-\lambda^a\cdot\psi_+(x^+)}
  A_{\{a\}_n}(x^+) ~,\qquad
  \psi^-_{\{a\}_n}(x^-)={\rm e}^{-\lambda^a\cdot\psi_-(x^-)}
  B_{\{a\}_n}(x^-) ~.
\end{eqnarray}
To simplify the expressions further we use condensed indices $A$, 
$\cdots$ for $\{a\}_n$, $\cdots$, and write (\ref{cexacsol2}) as 
\begin{eqnarray}
  \label{simp}
  {\rm e}^{-\lambda^a\cdot\varphi}=\sum_{A,B}C^a_{AB}\psi^+_A\star\psi^-_B~,
\end{eqnarray}
where we have included the cosmological constants in the numerical 
coefficients $C^a_{AB}$. The basic fact known for the chiral fields is 
that they satisfy quadratic Poisson algebras \cite{Babe,bdf,btb}
\begin{eqnarray}
  \label{qpoisalg}
  \{\psi^\pm_A(x),\psi^\pm_B(x')\}=-\frac{\gamma^2}{4}
  \sum_{C,D}[\theta(x-x')r_{AB}^{CD}
  -\theta(x'-x)\bar r_{AB}^{CD}]\psi^\pm_C(x)\psi^\pm_D(x')~,
\end{eqnarray}
where we have restricted ourselves to $0\le x,~x'<2\pi$ and $\theta(x)$ 
is the unit step function. The $r$-matrix may depend on 
the zero-mode momenta $P$ and must satisfy 
\begin{eqnarray}
  \label{rbar}
  \bar r_{AB}^{CD}=r_{BA}^{DC} ~,
\end{eqnarray}
due to the anti-symmetry property of the Poisson bracket.
They are also constrained by the classical Yang-Baxter 
equations. It is now straightforward to write down the condition for 
the canonicity of the transformation $\psi\rightarrow\varphi$ in terms 
of the $r$-matrix. We first consider the classical locality
\begin{eqnarray}
  \label{cloc}
  0\hskip -.2cm&=&\hskip -.2cm\{{\rm e}^{-\lambda^a\cdot\varphi(0,\sigma)},
  {\rm e}^{-\lambda^b\cdot\varphi(0,\sigma')}\} \nonumber\\
  \hskip -.2cm&=&\hskip -.2cm -\frac{\gamma^2}{4}
  \sum_{A,B}\sum_{A',B'}\sum_{C,C'}\Bigl[\theta(\sigma-\sigma')
  \bigl(C^a_{CB}C^b_{C'B'}r_{CC'}^{AA'}-C^a_{AC}C^b_{A'C'}\bar r_{CC'}^{BB'}
  \bigr) \nonumber\\
  \hskip -.2cm&&\hskip -.2cm+\theta(\sigma'-\sigma)\bigl(
  C^a_{AC}C^b_{A'C'}r_{CC'}^{BB'}-C^a_{CB}C^b_{C'B'}
  \bar r_{CC'}^{AA'}\bigr)\Bigr]\psi^+_A(\sigma)\psi^+_{A'}(\sigma')\star
  \psi^-_B(-\sigma)\psi^-_{B'}(-\sigma') ~.
\end{eqnarray}
This leads to 
\begin{eqnarray}
  \label{cloc2}
  \sum_{C,C'}\bigl(C^a_{AC}C^b_{A'C'}r_{CC'}^{BB'}-C^a_{CB}C^b_{C'B'}
  r_{C'C}^{A'A}\bigr)=0~.
\end{eqnarray}
We next consider the Poisson bracket 
$\{{\rm e}^{-\lambda^a\cdot\varphi(0,\sigma)},
\partial_+{\rm e}^{-\lambda^b\cdot\varphi(0,\sigma')}\}$. Using 
(\ref{cexacsol2}) and (\ref{qpoisalg}), we arrive at 
\begin{eqnarray}
  \label{cfpblhs}
  \{{\rm e}^{-\lambda^a\cdot\varphi(0,\sigma)},
  \partial_+{\rm e}^{-\lambda^b\cdot\varphi(0,\sigma')}\}
  \hskip -.2cm
  &=&\hskip -.2cm
  \frac{\gamma^2}{4}\delta(\sigma-\sigma')\sum_{A,B}\sum_{A',B'}\sum_{C,C'}
  C^a_{CB}C^b_{C'B'}(r_{CC'}^{AA'}+r_{C'C}^{A'A}) \nonumber\\
  &&\hskip 2cm\times\psi^+_A(\sigma)\psi^+_{A'}(\sigma)\star
  \psi^-_B(-\sigma)\psi^-_{B'}(-\sigma) ~.
\end{eqnarray}
In the derivation, use has been made of (\ref{cloc2}). Note that the Poisson 
bracket is proportional to $\delta(\sigma-\sigma')$.  Such $\delta$-function 
can only arise from the Poisson bracket 
\begin{eqnarray}
  \label{cfpb}
  \{{\rm e}^{-\lambda^a\cdot\psi(0,\sigma)},
  \partial_+{\rm e}^{-\lambda^b\cdot\psi(0,\sigma')}\}=
  \frac{\gamma^2}{2}\lambda^a\cdot\lambda^b
  \delta(\sigma-\sigma')
  {\rm e}^{-(\lambda^a+\lambda^b)\cdot\psi(0,\sigma)}~. 
\end{eqnarray}
This leads to the second fundamental Poisson bracket
\begin{eqnarray}
  \label{2ndfpb}
  \{\varphi_k(0,\sigma),\pi^l_\varphi(0,\sigma')\}
  =\{\psi_k(0,\sigma),\pi^l_\psi(0,\sigma')\}
  =\delta_k^l\delta(\sigma-\sigma')~.
\end{eqnarray}
Note that the classical locality automatically guarantees (\ref{2ndfpb}). 
We will see that similar situation also 
occurs in quantum theory. We thus see that the transformation from the 
free field $\psi$ to the Toda field $\varphi$ is a canonical mapping 
if the $r$-matrix satisfies (\ref{cloc2}). 

Though (\ref{cloc2}) suffices for the canonicity of the free field, it is 
interesting to write down the conditions leading to (\ref{2ndfpb}) 
in term of the $r$-matrix. In doing this we note that the chiral fields 
(\ref{psipm}) may happen to satisfy quadratic identities 
\begin{eqnarray}
  \label{quadid}
  \sum_{A,B}f_r^{AB}\psi_A^\pm(x)\psi_{B}^\pm(x)&=&0~,
\end{eqnarray}
where $f_r^{AB}=f_r^{BA}$ may depend on $P$ and $r$ labels the identities. 
These can be dealt with via multiplier method. From (\ref{cfpblhs}), 
(\ref{quadid}) and the expansion  
\begin{eqnarray}
  \label{cfpbrhs}
  {\rm e}^{-(\lambda^a+\lambda^b)\cdot\varphi(0,\sigma)}
  =\sum_{A,B}\sum_{A',B'}C^a_{AB}C^b_{A'B'}
  \psi^+_A(\sigma)\psi^+_{A'}(\sigma)\star
  \psi^-_B(-\sigma)\psi^-_{B'}(-\sigma) ~,
\end{eqnarray}
we obtain the desired condition for the $r$-matrix
\begin{eqnarray}
  \label{cfpb2}
  &&\sum_{C,C'}(C^a_{AC}C^b_{A'C'}+C^a_{A'C}C^b_{AC'})
  (r_{CC'}^{BB'}+r_{CC'}^{B'B}+r_{C'C}^{B'B}+r_{C'C}^{BB'})
  +\sum_r(\mu_{AA'}^{ab;r}f_r^{BB'}+\nu_{BB'}^{ab;r}f_r^{AA'})
  \nonumber\\
  &&\hskip 2.cm=\rho^{ab}(C^a_{AB}C^b_{A'B'}+C^a_{A'B}C^b_{AB'}
  +C^a_{AB'}C^b_{A'B}+C^a_{A'B'}C^b_{AB}) ~,
\end{eqnarray}
where $\mu_{AA'}^{ab;r}$ and $\nu_{BB'}^{ab;r}$ are the multipliers 
for the quadratic identities (\ref{quadid}) and we have defined 
$\displaystyle{\rho^{ab}=2\lambda^a\cdot\lambda^b}$. 
The fundamental Poisson brackets can also be established by examining 
(\ref{cloc2}) and (\ref{cfpb2}). 

As an illustration, we consider the Liouville case. Though the complete 
algebra is contained in that of $A_2$-Toda theory, we examine (\ref{cloc2}) 
and (\ref{cfpb2}) for the Liouville case, separately. Choosing the only 
simple root to be unity, we define the chiral fields by 
\begin{eqnarray}
  \label{liov}
  \psi_0^+(x^+)={\rm e}^{-\frac{1}{2}\psi_+(x^+)}~, \qquad
  \psi^+_1(x^+)={\rm e}^{-\frac{1}{2}\psi_+(x^+)}A(x^+) ~,
\end{eqnarray}
where $A(x)$ is assumed to satisfy $\partial_+ A={\rm e}^{\psi_+}$ as in 
(\ref{aseq}), and the right-moving fields are similarly defined. 
The nonvanishing components of the $r$-matrix are given by
\begin{eqnarray}
  \label{liouvrmat}
  r_{00}^{00}=-r_{01}^{01}=-r_{10}^{10}=r_{11}^{11}=\frac{1}{4} ~, \quad
  r_{01}^{10}=\frac{1}{2}\Bigl(1+\coth\frac{\gamma}{2}P\Bigr)~, 
  \quad r_{10}^{01}=\frac{1}{2}\Bigl(1-\coth\frac{\gamma}{2}P\Bigr)~.
\end{eqnarray}
In the Liouville case there is no quadratic identity of the form 
(\ref{quadid}) and the conditions (\ref{cloc2}) and (\ref{cfpb2}) are 
reduced to
\begin{eqnarray}
  \label{lcons}
  r_{AA'}^{BB'}=r_{B'B}^{A'A} ~, \qquad 
  r_{AA'}^{BB'}+r_{AA'}^{B'B}+r_{A'A}^{B'B}+r_{A'A}^{BB'}
  =\frac{1}{2}(\delta_{AB}\delta_{A'B'}+\delta_{AB'}\delta_{A'B})~.
\end{eqnarray}
These relations are obviously satisfied by the $r$-matrix (\ref{liouvrmat}). 

We now turn to the case of $A_2$-Toda theory. We use the convention 
$(\alpha^a)^2=2$ for the simple roots and define the chiral fields by
\begin{eqnarray}
  \label{a2cf}
  \psi^+_0(x^+)={\rm e}^{-\lambda^1\cdot\psi_+(x^+)} , ~
  \psi^+_1(x^+)={\rm e}^{-\lambda^1\cdot\psi_+(x^+)}A_1(x^+) , ~
  \psi^+_2(x^+)={\rm e}^{-\lambda^1\cdot\psi_+(x^+)}A_{12}(x^+) ,
  \nonumber \\
  \psi^+_{\bar 0}(x^+)={\rm e}^{-\lambda^2\cdot\psi_+(x^+)} , ~
  \psi^+_{\bar 1}(x^+)={\rm e}^{-\lambda^2\cdot\psi_+(x^+)}A_2(x^+) , ~
  \psi^+_{\bar 2}(x^+)={\rm e}^{-\lambda^2\cdot\psi_+(x^+)}A_{21}(x^+) .
\end{eqnarray}
The $r$-matrix satisfies $r_{A\bar A}^{\bar BB}=r_{\bar AA}^{B\bar B}=0$,  
and the charge conservation $r_{AB}^{CD}=r_{A\bar B}^{C\bar D}=0$ unless 
$A+B=C+D$. Then the nonvanishing elements can be easily read off from the 
quadratic Poisson algebra given in Appendix \ref{sec:appA} as 
\begin{eqnarray}
  \label{a2rmat}
  && r_{00}^{00}=r_{11}^{11}=r_{22}^{22}=\frac{2}{3}~, \quad 
  r_{01}^{01}=r_{10}^{10}=r_{02}^{02}=r_{20}^{20}=r_{12}^{12}=r_{21}^{21}
  =-\frac{1}{3} ~,\nonumber \\
  &&r_{01}^{10}=1-\coth\frac{\gamma}{4}\alpha^a\cdot P ~, \quad
  r_{10}^{01}=1+\coth\frac{\gamma}{4}\alpha^a\cdot P ~,\nonumber\\
  &&r_{02}^{20}=1-\coth\frac{\gamma}{4}(\alpha^a+\alpha^{\bar a})\cdot P ~, 
  \quad
  r_{20}^{02}=1+\coth\frac{\gamma}{4}(\alpha^a+\alpha^{\bar a})\cdot P ~,
  \nonumber \\
  &&r_{12}^{21}=1-\coth\frac{\gamma}{4}\alpha^{\bar a}\cdot P ~, \quad
  r_{21}^{12}=1+\coth\frac{\gamma}{4}\alpha^{\bar a}\cdot P ~, \nonumber\\
  &&r_{0\bar 0}^{0\bar 0}=r_{\bar 00}^{\bar 00}
  =r_{0\bar1}^{0\bar1}=r_{\bar01}^{\bar01}
  =r_{1\bar0}^{1\bar0}=r_{\bar10}^{\bar10} 
  =r_{1\bar2}^{1\bar2}=r_{\bar12}^{\bar12}
  =r_{2\bar1}^{2\bar1}=r_{\bar21}^{\bar21}
  =r_{2\bar2}^{2\bar2}=r_{\bar22}^{\bar22}=\frac{1}{3} ~,\nonumber\\
  &&r_{0\bar2}^{0\bar2}=r_{\bar02}^{\bar02}
  =r_{2\bar0}^{2\bar0}=r_{\bar20}^{\bar20}
  =r_{1\bar1}^{1\bar1}=r_{\bar11}^{\bar11}=-\frac{2}{3} ~,\nonumber\\
  &&r_{0\bar2}^{1\bar1}=r_{\bar11}^{\bar20}
  =1-\coth\frac{\gamma}{4}\alpha^a\cdot P ~, \quad
  r_{1\bar1}^{0\bar2}
  =r_{\bar20}^{\bar11}=1+\coth\frac{\gamma}{4}\alpha^a\cdot P ~, \nonumber\\
  &&r_{1\bar1}^{2\bar0}=r_{\bar02}^{\bar11}
  =1-\coth\frac{\gamma}{4}\alpha^{\bar a}\cdot P ~, \quad
  r_{2\bar0}^{1\bar1}
  =r_{\bar11}^{\bar02}=1+\coth\frac{\gamma}{4}\alpha^{\bar a}\cdot P ~, 
  \nonumber\\
  &&r_{0\bar2}^{2\bar0}=r_{\bar02}^{\bar20}=-1+\coth\frac{\gamma}{4}
  (\alpha^a+\alpha^{\bar a})\cdot P ~,\quad 
  r_{2\bar0}^{0\bar2}=r_{\bar20}^{\bar02}=-1-\coth\frac{\gamma}{4}
  (\alpha^a+\alpha^{\bar a})\cdot P ~.\nonumber\\
\end{eqnarray}
In the present case there is a pair of quadratic identities
\begin{eqnarray}
  \label{a2qid}
  \psi_1^\pm(x)\psi_{\bar1}^\pm(x)-\psi_0^\pm(x)\psi_{\bar2}^\pm(x)
  -\psi_2^\pm(x)\psi_{\bar0}^\pm(x)=0~.
\end{eqnarray}
The conditions to be examined out of (\ref{cloc2}) and (\ref{cfpb2}) 
are then given by
\begin{eqnarray}
  \label{a2cons}
  && r_{AA'}^{BB'}=r_{B'B}^{A'A} ~, \quad
  r_{AA'}^{BB'}+r_{AA'}^{B'B}+r_{A'A}^{B'B}+r_{A'A}^{BB'}
  =\frac{4}{3}(\delta_{AB}\delta_{A'B'}+\delta_{AB'}\delta_{A'B})~,
  \nonumber\\
  &&r_{A\bar A}^{B\bar B}=r_{\bar BB}^{\bar AA} ~, \quad
  r_{A\bar A}^{B\bar B}+r_{\bar AA}^{\bar BB}
  +\mu_{A\bar A}f^{B\bar B}+\nu_{B\bar B}f^{A\bar A}
  =\frac{2}{3}\delta_{AB}\delta_{\bar A\bar B}~,  
\end{eqnarray}
where the nonvanishing coefficients arising from (\ref{a2qid}) are 
$f^{1\bar1}=-f^{0\bar2}=-f^{2\bar0}=1$. The multipliers $\mu_{A\bar A}$ 
and $\nu_{B\bar B}$ are not uniquely determined since the simultaneous shifts 
$\mu_{A\bar A}\rightarrow\mu_{A\bar A}+\kappa f^{A\bar A}$, 
$\nu_{B\bar B}\rightarrow\nu_{B\bar B}-\kappa f^{B\bar B}$ for arbitrary 
$\kappa$ has no effect on (\ref{a2cons}). One can directly verify that 
the $r$-matrix (\ref{a2rmat}) 
satisfies (\ref{a2cons}) for the multipliers $\mu_{A\bar A}=\nu_{A\bar A}
=-f^{A\bar A}$. This establishes that the transformation from the 
free fields to the Toda fields defined by the classical solution
(\ref{a2csol}) is indeed a canonical transformation.

\section{Quantum $A_2$-Toda Theory}
\label{sec:qa2toda}
\setcounter{equation}{0}

After a somewhat long preliminary description of the chiral structure 
of classical Toda field theories from the canonical theoretic view point, we 
turn to the quantization of $A_2$-Toda field. The reason of the 
restriction of our analysis to the specific case is that it is 
the simplest but nontrivial extension of Liouville theory and can be 
expected to reveal common features of higher rank theories, for 
which direct canonical approach presented in this paper might be 
difficult to apply. 

As we argued in the previous section, the interacting Toda fields 
can be expressed in terms of the canonical free fields. Then the 
quantization of the system can be carried out by imposing the canonical 
commutation relations for the free fields by the prescription 
$\{~,~\}\rightarrow\frac{1}{i}[~,~]$. Equivalently, we may assume that 
the normal modes satisfy the standard commutation relations
\begin{eqnarray}
  \label{ccrpsi}
  [Q_k,P_l]=i\delta_{kl}, \qquad
  [a^{(+)}_{kn},a^{(+)}_{lm}]
  =[a^{(-)}_{kn},a^{(-)}_{lm}]=n\delta_{n+m,0}\delta_{kl}~.\quad 
  (k,l=1,2)
\end{eqnarray}
The Hilbert space of the state vectors of the Toda field sector 
is the direct sum $\bigoplus_{p}{\cal H}_{p}$, where
$p$ specifies the eigenvalue of $P$ and ${\cal H}_{p}$ stands 
for the Fock space generated by the oscillators $a^{(\pm)}_{kn}$ ($n<0$) 
from the ground state with eigenvalue $p$. The Toda exponential operators, 
however, do not have a well-defined action on the Hilbert space since they 
are composed of operators shifting $p$ to imaginary direction in a formal 
sense. We do not consider such subtleties in this article. See the 
remarks of sect. 4.1.6 of ref. \cite{gs94}. 

The chiral vertex operators are defined by the free field normal ordering
for the oscillatory modes and the symmetric ordering for the zero-mode 
operators by the rule $:{\rm e}^{\beta\cdot Q}f(P):={\rm e}^{\frac{1}{2}
\beta\cdot Q}f(P){\rm e}^{\frac{1}{2}\beta\cdot Q}$. We redefine the chiral 
screening charges $A_a$ and $A_{a\bar a}$ by 
\begin{eqnarray}
  \label{cscop}
  A_a(x)&=&\int_0^{2\pi}dy:{\cal E}_{\alpha^a}(x-y)V^+_a(y): ~,
  \nonumber\\
  A_{a\bar a}(x)&=&\int_0^{2\pi}dydz
  :{\cal E}_{\alpha^a}(x-y)V^+_a(y):
  :{\cal E}_{\alpha^{\bar a}}(x-y)
  {\cal E}_{\alpha^{\bar a}}(y-z)V^+_{\bar a}(z):~. 
\end{eqnarray}
Two major modifications are made here from the 
classical expressions (\ref{a}). The one is the omission of the overall 
$P$ dependent coefficients to simplify the operator algebra satisfied 
by the screening charges. The other is the rescaling $\psi_+\rightarrow
\eta\psi_+$ to keep the conformal weight of the vertex operator $V^+_a$ 
to be $(1,0)$ \cite{ct,gn84,ow}. We assume that the conformal 
symmetry is 
generated by the normal-ordered free field stress tensor given by
\begin{eqnarray}
  \label{qst}
  T_{\pm\pm}&=&\frac{1}{\gamma^2}
  (:\partial_\pm\psi\cdot\partial_\pm\psi:
  -2\rho\cdot\partial_\pm^2\psi) ~.
\end{eqnarray}
Then the conformal weight $\Delta_\beta$ of the vertex operators
$:{\rm e}^{\eta\beta\cdot\psi}:$ is given by 
\begin{eqnarray}
  \label{cdim}
  \Delta_\beta=\beta\cdot\Bigl(\eta\rho-\frac{\gamma^2\eta^2}{8\pi}
  \beta\Biggr)~.
\end{eqnarray}
In particular $\eta$ satisfies 
\begin{eqnarray}
  \label{ita}
  \Delta_{\alpha^a}=\eta-\frac{\gamma^2\eta^2}{4\pi}=1 ~.
\end{eqnarray}
We also make the replacement  $P$ by $\eta P$ in 
${\cal E}_\beta(x)$ to ensure the quasi-periodicity of the screening 
charges under the shift $x\rightarrow x+2\pi$. 
The screening charges $B_a$ and $B_{a\bar a}$ in the right-moving sector 
are similarly defined. 

For later convenience, we introduce here some 
notation
\begin{eqnarray}
  \label{somedef}
  \varpi\equiv-\frac{iP}{\gamma\eta}~, \qquad
  \varpi^a\equiv\alpha^a\cdot\varpi ~, \qquad
  g\equiv\frac{\gamma^2\eta^2}{8\pi}=\frac{\eta-1}{2}~, \qquad
  q\equiv{\rm e}^{2\pi ig}~.
\end{eqnarray}
We will see that $q$ is the quantum deformation parameter and, hence, 
$g$ plays the role of the Planck constant. In terms of these 
variables (\ref{cscop}) can be written as
\begin{eqnarray}
  \label{cscop2}
  A_a(x)&=&\int_0^{2\pi}dyq^{(\varpi^a+1)\epsilon(x-y)}
  V_a^+(y) \nonumber\\
  A_{a\bar a}(x)&=&\int_0^{2\pi}dydz
  q^{(\varpi^a+\varpi^{\bar a}+1)\epsilon(x-y)
    +\varpi^{\bar a}\epsilon(y-z)} V_a^+(y)V^+_{\bar a}(z)~.
\end{eqnarray}
In quantum theory the quasi-periodicity (\ref{quasip}) is modified to
\begin{eqnarray}
  \label{qquasip}
  A_a(x+2\pi)=q^{2(\varpi^a+1)}A_a(x) ~, \quad
  A_{a\bar a}(x+2\pi)=q^{2(\varpi^a+\varpi^{\bar a}+1)}A_{a\bar a}(x) ~
\end{eqnarray}
The screening charges and the vertex operators $V^{a\pm}_\kappa
\equiv:{\rm e}^{\eta\kappa\lambda^a\cdot\psi_\pm}:$ are the building 
blocks of the Toda exponential operators. They are hermitian for the standard 
assignment of hermiticity $\psi_\pm^\dagger=\psi_\pm$ as can be shown by 
noting the relation
\begin{eqnarray}
  \label{VV}
  :{\rm e}^{\eta\beta\cdot\psi_\pm(x)}::{\rm e}^{\eta\beta'\cdot\psi_\pm(y)}:
  =q^{-\beta\cdot\beta'\epsilon(x-y)}
  :{\rm e}^{\eta\beta'\cdot\psi_\pm(y)}:
  :{\rm e}^{\eta\beta\cdot\psi_\pm(x)}: ~.
\end{eqnarray}
Furthermore they satisfy a characteristic exchange algebra. In Appendix 
\ref{sec:appB} we summarize the exchange properties of the screening 
charges needed for the computation of the ${\cal R}$-matrix discussed 
below. In the construction of the Toda exponential operator the crucial 
property of the chiral vertex operators is the mutual commutativity of 
$V^{a+}_\kappa(x)$, $A_a(x)$ and $A_{a\bar a}(x)$. 

The $\star$-product (\ref{starp}) introduced in the previous section 
can be defined also in quantum theory. Let $L$ and $R$, respectively, 
be the left- and the right-chiral operators satisfying 
$[L{\rm e}^{-\gamma\eta\omega\cdot Q},{\rm e}^{-\gamma\eta\omega\cdot Q}R]
=0$  for some numerical constant $\omega$, then we define their 
$\star$-product by
\begin{eqnarray}
  \label{qstarp}
  L\star R\equiv L{\rm e}^{-\gamma\eta\omega\cdot Q}R ~.
\end{eqnarray}
The $\star$-product satisfies the multiplication rule
\begin{eqnarray}
  \label{qstp}
  L_1\star R_1 L_2\star R_2=L_1L_2\star R_1R_2~. 
\end{eqnarray}
This is the quantum generalization of (\ref{pbialg}). It implies 
that the left- and the right-chiral operators can be considered as 
commuting under the $\star$-product. In particular one easily see 
$L\star f(\varpi)R=f(\varpi)L\star R$ for arbitrary $f$ as a special 
case of (\ref{qstp}). 

In terms of the $\star$-product the screening charges ${\cal Y}_a$ 
and ${\cal Y}_{a\bar a}$ introduced in ref. \cite{fit98} are simply given by
\begin{eqnarray}
  \label{yayaab}
  {\cal Y}_a(x)=A_a(x^+)\star B_a(x^-)~, \qquad
  {\cal Y}_{a\bar a}(x)=A_{a\bar a}(x^+)\star B_{a\bar a}(x^-)~.
\end{eqnarray}
Then the Toda exponential operator can be expressed as a power series 
of the screening charges 
\begin{eqnarray}
  \label{teo}
  {\rm e}^{\eta\kappa\lambda^a\cdot\varphi(x)}
  &=&V_\kappa^a(x)\sum_{n,m=0}^{\infty}\Biggl(\frac{\mu^2}{4}\Biggr)^{n+2m}
  C^a_{nm}(\kappa;\varpi)({\cal Y}_a(x))^n({\cal Y}_{a\bar a}(x))^m 
  \nonumber\\
  &=&\sum_{n,m=0}^{\infty}\Biggl(\frac{\mu^2}{4}\Biggr)^{n+2m} 
  C^a_{nm}(\kappa;\varpi+\kappa\lambda^a)\psi^{a+}_{nm}(\kappa;x^+)
  \star \psi^{a-}_{nm}(\kappa;x^-)
\end{eqnarray}
where  $V^a_\kappa(x)=V^{a+}_\kappa(x^+)\star V^{a-}_\kappa(x^-)$ 
is the free field vertex operator in the vector scheme. 
We have rescaled $\varphi$ by $\eta$ as was done for the canonical 
free field to make the operatorial mapping $\psi\rightarrow\varphi$ 
to keep the canonical commutation relations.\footnote{In the previous 
work \cite{fit98} this rescaling was not considered for the Toda 
fields.} We have also introduced the chiral fields by
\begin{eqnarray}
  \label{qchif}
  \psi^{a+}_{nm}(\kappa;x)=V^{a+}_\kappa(x)(A_a(x))^n
  (A_{a\bar a}(x))^m~, \quad
  \psi^{a-}_{nm}(\kappa;x)=V^{a-}_\kappa(x)(B_a(x))^n
  (B_{a\bar a}(x))^m ~.
\end{eqnarray}
Since the vertex operators and the screening charges are commuting, 
there arises no ordering problem in the expansion (\ref{teo}). 
The conformal covariance of the Toda exponential is attributed to 
that of the free field vertex operator. The coefficients $C^a_{nm}
(\kappa;\varpi)$ may depend on the zero-mode $\varpi$ without 
affecting the conformal symmetry \cite{ow}. They are assumed to satisfy
the conditions
\begin{eqnarray}
  \label{Cscon}
  &&C_{00}^a(\kappa;\varpi)=1 ~, \quad C^a_{nm}(0;\varpi)=\delta_{n0}
  \delta_{m0} ~, \nonumber\\
  && (C^a_{nm}(\kappa;\varpi))^\dagger=C^a_{nm}(\kappa;\varpi+\kappa\lambda^a
  -(n+m)\alpha^a-m\alpha^{\bar a})~.
\end{eqnarray}
The last relation corresponds to the hermiticity of the Toda exponential 
operator for real $\kappa$. In ref. \cite{fit98} the coefficients $C^a_{nm}$ 
are explicitly given as a conjecture based upon the analysis of 
the locality condition
\begin{eqnarray}
  \label{qloc}
  [e^{\kappa\lambda^a\cdot\varphi(0,\sigma)},
e^{\kappa\lambda^b\cdot\varphi(0,\sigma')}]=0 
\end{eqnarray}
up to the fourth 
order in $\mu^2$. We will fill the gap and give a complete 
proof for the conjectured forms of the coefficients in the next 
section. 

At this point we introduce the additive $\varpi$-charges. If an operator 
${\cal O}$ satisfies $[\varpi^a,{\cal O}]=-\nu^a{\cal O}$, we 
assign ${\cal O}$ the $\varpi^a$-charges $\nu^a$. Then the chiral 
fields (\ref{qchif}) can be assigned definite charges 
since the $\varpi^a$-charges of the vertex operators $V_\kappa^{a\pm}$ 
and $V_a^\pm$  are, respectively, $\kappa$ and $2$, whereas the 
$\varpi^{\bar a}$-charges are $0$ and $-1$.

One can easily be convinced from the operator algebra given in 
Appendix \ref{sec:appB} that the chiral fields form a closed 
exchange algebra in each chiral sector as 
\begin{eqnarray}
  \label{excalg}
  \psi^{a\pm}_{nm}(\kappa;x)\psi^{b\pm}_{n'm'}(\nu;x')
  &=&\cases{\displaystyle{\sum_{r,r',s,s'}
  {\cal R}_{nm}^{rs}{}_{;n'm'}^{;r's'}({\scriptstyle {a\atop 
      \kappa}}|{\scriptstyle {b\atop \nu}};\varpi\bigr)
  \psi^{b\pm}_{r's'}(\nu;x')
  \psi^{a\pm}_{rs}(\kappa;x)} &$(x>x')$\cr
  \displaystyle{\sum_{r,r',s,s}
  \overline{\cal R}_{nm}^{rs}{}_{;n'm'}^{;r's'}(
  {\scriptstyle {a\atop\kappa}}|{\scriptstyle {b\atop \nu}};\varpi\bigr)
  \psi^{b\pm}_{r's'}(\nu;x')
  \psi^{a\pm}_{rs}(\kappa;x)} &$(x<x')$\cr}
\end{eqnarray}
where the ${\cal R}$-matrix depends on the zero-mode momentum 
$\varpi$ and the sum should be taken over nonnegative integers $r$, $r'$, 
$s$, $s'$ satisfying $r+r'=n+n'$, $s+s'=m+m'$ if $b=a$ and 
$r+s+s'=n+m+m'$, $r'+s+s'=n'+m+m'$ if $b=\bar a$ due to the 
charge conservation. We can implement the charge conservation 
by assuming 
\begin{eqnarray}
  \label{cc}
  &&{\cal R}_{nm}^{rs}{}_{;n'm'}^{;r's'}({\scriptstyle {a\atop 
      \kappa}}|{\scriptstyle {b\atop \nu}};\varpi\bigr)
  =\overline{\cal R}_{nm}^{rs}{}_{;n'm'}^{;r's'}(
  {\scriptstyle {a\atop\kappa}}|{\scriptstyle {b\atop \nu}};
  \varpi\bigr)=0 \nonumber\\ 
  && {\rm unless} \quad
  \cases{r+r'=n+n',~s+s'=m+m' &for $b=a$\cr
    r+s+s'=n+m+m', ~r'+s+s'=n'+m+m' &for $b=\bar a$}
\end{eqnarray}

Applying the exchange algebra twice
for $\psi^{a+}_{kl}(\kappa;x)\psi^{b+}_{nm}(\nu;x')$ ($x<x'$) is simply 
the identity operation. This implies the consistency 
conditions 
\begin{eqnarray}
  \label{cRcond}
  \sum_{k',l',n',m'}
  \overline{\cal R}_{kl}^{k'l'}{}_{;nm}^{;n'm'}({\scriptstyle{a\atop 
      \kappa}}|{\scriptstyle {b\atop \nu}};\varpi\bigr)
  {\cal R}_{n'm'}^{n''m''}{}_{;k'l'}^{;k''l''}
  ({\scriptstyle{b\atop 
      \nu}}|{\scriptstyle {a\atop \kappa}};\varpi\bigr)=\delta_k^{k''}
  \delta_l^{l''}\delta_n^{n''}\delta_m^{m''} ~.
\end{eqnarray}
Hence, $\overline{\cal R}$ can be considered as the inverse of ${\cal R}$. 
Furthermore, the associativity of the operator 
products leads to the Yang-Baxter relations \cite{gn84,bg89}. For example, 
consider the two different ways of applying the exchange 
algebra for 
\begin{eqnarray}
  \label{opasoc}
  (\psi^{a+}_{kl}(\kappa;x)
  \psi^{b+}_{nm}(\nu;x'))\psi^{c+}_{rs}(\rho;x'')
  =\psi^{a+}_{kl}(\kappa;x)
  (\psi^{b+}_{nm}(\nu;x')\psi^{c+}_{rs}(\rho;x''))~.
  \quad (x>x'>x'')
\end{eqnarray}
Such operations are consistent only if the ${\cal R}$-matrix 
satisfies the relations
\begin{eqnarray}
  \label{YBeq}
  &&\sum_{k',l',n',m',r',s'}{\cal R}_{kl}^{k'l'}{}_{;nm}^{;n'm'}
  ({\scriptstyle{a\atop 
      \kappa}}|{\scriptstyle {b\atop \nu}};\varpi\bigr)
  {\cal R}_{k'l'}^{k''l''}{}_{;rs}^{;r's'}({\scriptstyle{a\atop 
      \kappa}}|{\scriptstyle {c\atop \rho}};\varpi+\beta_{n'm'}^{b;\nu}\bigr)
  {\cal R}_{n'm'}^{n''m''}{}_{;r's'}^{;r''s''}({\scriptstyle{b\atop 
      \nu}}|{\scriptstyle {c\atop \rho}};\varpi\bigr) \nonumber\\
  &&\hskip .5cm =\sum_{k',l',n',m',r',s'}{\cal R}_{kl}^{k'l'}
  {}_{;r's'}^{;r''s''}({\scriptstyle{a\atop 
      \kappa}}|{\scriptstyle {c\atop \rho}};\varpi\bigr)
  {\cal R}_{k'l'}^{k''l''}{}_{;n'm'}^{;n''m''}
  ({\scriptstyle{a\atop 
      \kappa}}|{\scriptstyle {b\atop \nu}};
  \varpi+\beta_{r''s''}^{c;\rho}\bigr)
  {\cal R}_{mn}^{n'm'}{}_{;rs}^{;r's'}
  ({\scriptstyle{b\atop 
      \nu}}|{\scriptstyle {c\atop \rho}};
  \varpi+\beta_{kl}^{a;\kappa}\bigr) ~,\nonumber\\
\end{eqnarray}
where $\beta^{a;\kappa}_{nm}$ stands for the shift of $\varpi$ arising 
in moving it from the right of $\psi^a_{nm}(\kappa;x)$ to the left and is 
given by
\begin{eqnarray}
  \label{beta}
  \beta^{a;\kappa}_{nm}=\kappa\lambda^a+(n+m)\alpha^a+m\alpha^{\bar a}~.
\end{eqnarray}
There are similar relations corresponding to different choices of 
the operator products. 

We now turn to the locality (\ref{qloc}). Using (\ref{teo}) and 
(\ref{excalg}), we can rewrite it as a condition for the ${\cal R}$-matrix
as 
\begin{eqnarray}
  \label{locR}
  &&\sum_{n,n',m,m'}C^a_{nm}(\kappa;\varpi
  +\kappa\lambda^a)C^b_{n'm'}(\nu;\varpi
  +\nu\lambda^b+\beta^{a;\kappa}_{nm}){\cal R}_{nm}^{kl}
  {}_{;n'm'}^{;k'l'}
  ({\scriptstyle {a\atop \kappa}}|{\scriptstyle {b\atop \nu}};\varpi\bigr)
  \overline{\cal R}_{nm}^{rs}{}_{;n'm'}^{;r's'}(
  {\scriptstyle {a\atop\kappa}}|{\scriptstyle {b\atop \nu}};\varpi\bigr)
  \nonumber\\
  &&\hskip 3cm=\delta_{kr}\delta_{ls}\delta_{k'r'}\delta_{l's'}
  C^a_{kl}(\kappa;\varpi+\kappa\lambda^a+\beta^{b;\nu}_{k'l'})
  C^b_{k'l'}(\nu;\varpi+\nu\lambda^b) ~.
\end{eqnarray}
By noting (\ref{cRcond}), one can alternatively express this 
in a simpler form as  
\begin{eqnarray}
  \label{locR2}
  &&C^a_{nm}(\kappa;\varpi
  +\kappa\lambda^a)C^b_{n'm'}(\nu;\varpi
  +\nu\lambda^b+\beta^{a;\kappa}_{nm})
  {\cal R}_{nm}^{rs}{}_{;n'm'}^{;r's'}
  ({\scriptstyle {a\atop\kappa}}|{\scriptstyle {b\atop \nu}};\varpi\bigr)
  \nonumber\\
  &&\hskip 2cm=C^a_{rs}(\kappa;\varpi+\kappa\lambda^a
  +\beta^{b;\nu}_{r's'})
  C^b_{r's'}(\nu;\varpi+\nu\lambda^b)
  {\cal R}_{r's'}^{n'm'}{}_{;rs}^{;nm}
  ({\scriptstyle {b\atop\nu}}|{\scriptstyle {a\atop\kappa}};\varpi\bigr) ~.
\end{eqnarray}
We see that the locality of the quantum Toda fields can be 
attributed to the property of the ${\cal R}$-matrix as in the classical 
theory and (\ref{locR2}) corresponds to the quantum theoretical extension 
of the condition (\ref{cloc2}) for the $A_2$-Toda theory. 

The quantum Toda field $\varphi$ can be obtained from the Toda exponential 
operator (\ref{teo}) as the derivative with respect to $\kappa$ at the 
origin $\kappa=0$. This leads to
\begin{eqnarray}
  \label{qtf}
  \eta\lambda^a\cdot\varphi(x)=\eta\lambda^a\cdot\psi(x)
  +\Upsilon^a(x) ~,
\end{eqnarray}
where $\Upsilon^a(x)$ is a power series of the screening charges and will be 
given in sect. \ref{sec:feq} in establishing the field equations. We 
need not here the explicit form. Using (\ref{locR}), one can show 
for $\sigma\ne\sigma'$
\begin{eqnarray}
  \label{etc2}
  [{\rm e}^{\eta\kappa\lambda^a\cdot\varphi(0,\sigma)},
\partial_\tau{\rm e}^{\eta\nu\lambda^a\cdot\varphi(0,\sigma')}]
=[\partial_\tau{\rm e}^{\eta\kappa\lambda^a\cdot\varphi(0,\sigma)},
\partial_\tau{\rm e}^{\eta\nu\lambda^a\cdot\varphi(0,\sigma')}]=0 ~. 
\end{eqnarray}
This together with (\ref{qloc}) imply the 
following relations 
\begin{eqnarray}
  \label{etc3}
  &&[\eta\lambda^a\cdot\psi(0,\sigma),\Upsilon^b(0,\sigma')]
  +[\Upsilon^a(0,\sigma),\eta\lambda^b\cdot\psi(0,\sigma')]
  +[\Upsilon^a(0,\sigma),\Upsilon^b(0,\sigma')]=0~, \nonumber\\
  &&[\eta\lambda^a\cdot\psi(0,\sigma),\dot\Upsilon^b(0,\sigma')]
  +[\Upsilon^a(0,\sigma),\eta\lambda^b\cdot\dot\psi(0,\sigma')]
  +[\Upsilon^a(0,\sigma),\dot\Upsilon^b(0,\sigma')]=0~,\\ 
  &&[\eta\lambda^a\cdot\dot\psi(0,\sigma),\dot\Upsilon^b(0,\sigma')]
  +[\dot\Upsilon^a(0,\sigma),\eta\lambda^b\cdot\dot\psi(0,\sigma')]
  +[\dot\Upsilon^a(0,\sigma),\dot\Upsilon^b(0,\sigma')]=0~.\nonumber
\end{eqnarray}
The lhs' of these expressions may exhibit at most finite discontinuities 
at $\sigma=\sigma'$. Hence the contributions from 
$\Upsilon^a$ are canceled in the equal-time commutation 
relations between $\varphi$ and $\displaystyle{\pi_\varphi
\equiv\frac{1}{\gamma^2}\dot\varphi}$. We thus arrive at the full set 
of canonical commutation relations 
\begin{eqnarray}
  \label{etcc}
  &&[\varphi_k(0,\sigma),\varphi_l(0,\sigma')]
  =[\pi^k_\varphi(0,\sigma),\pi^l_\varphi(0,\sigma')]=0~, 
  \nonumber\\
  &&[\varphi_k(0,\sigma),\pi^l_\varphi(0,\sigma')]
  =[\psi_k(0,\sigma),\pi^l_\psi(0,\sigma')]
  =i\delta_k^l\delta(\sigma-\sigma')~.
\end{eqnarray}
As in the classical theory, the locality of the Toda exponential 
operators ensures the set of canonical commutation relations of the
Toda fields. This establishes that the operatorial transformation 
(\ref{teo}) from $\psi$ to $\varphi$ induces a canonical mapping 
between the canonical pairs of the Toda system and those of the free 
theory. 

In ref. \cite{fit96} the locality and the canonical commutation 
relations in Liouville theory were separately discussed and their 
intimate relationship was not so clarified. This was mainly due to 
the use of the vector scheme. As has been thoroughly investigated 
for Liouville theory in refs. \cite{gn,gs93}, the virtue of the chiral 
description is that it not only enables systematic analysis of the 
locality but also makes clear the role of the underlying quantum group 
symmetry of the operator algebra of the chiral fields. 

Before closing the present section we make a remark for the 
generalization to higher rank cases. In general only a specific 
set of screening charges appears in the expansion of the Toda 
exponential operator associated with an individual fundamental 
weight. If such screening charges and the free field vertex 
operator associated with the corresponding fundamental weight 
are mutually commuting as in the $A_1$ and $A_2$ cases, one 
can define chiral fields without the ordering problem. Since 
the chiral fields are expected to form a closed exchange 
algebra, the above arguments for the $A_2$ case are considered to be 
applicable for general Toda theories without essential modifications.

\section{Quantum $A_2$-Toda Exponential Operators}
\label{sec:qteo}
\setcounter{equation}{0}

As announced in the previous section we determine the coefficients 
$C^a_{nm}(\kappa;\varpi)$ by requiring the locality (\ref{qloc}) for 
the Toda exponential operator (\ref{teo}). In ref. \cite{fit98} the locality 
is investigated directly order by order in the cosmological constant 
$\mu^2$. This approach suffices to guess the general forms of the 
expansion coefficients. However, such direct method does not seem to 
work well due to the increasing complication in handling the operator 
equations if one tries to find and solve the constraints on the 
coefficients at an arbitrary higher order. In order to systematically 
analyze the locality constraints it is desired to extend the approach 
of ref. \cite{fit96} developed for Liouville theory to the $A_2$ case. 
The point that makes it possible to solve the locality conditions 
in the Liouville case is that the screening charge can be decomposed 
into a set of operators satisfing simple exchange algebra which allows 
one dimensional quantum mechanical realization. Making use of the 
property, one can convert the locality condition in operator form 
into the algebraic relation containing only commuting variables. 
It integrates the functional recurrence relations for the expansion 
coefficients of the Liouville exponential operator with two arbitrary 
parameters and can be solved with respect to the expansion 
coefficients by making use of the arbitrariness of the two parameters. 
This successful approach  can be applied also for the $A_2$-Toda theory
with some modifications. The sharp difference with the $A_1$ case 
is that the screening charges (\ref{cscop2}) can not be decomposed 
into a set of operators which would lead to the generalization of the 
integrated recurrence relations for the $A_1$ case through 
a finite dimensional quamtum mechanical realization. Fortunately, 
we do not need to take acount of the full operator algebra of the 
screening charges as is given in Appendix \ref{sec:appB}. We see 
that a judicious choice of a set of the independent components of 
operators appearing in the screening charges that allow a quantum 
mechanical realization is sufficient for the purpose to determine 
the expansion coefficients. 

The locality condition (\ref{qloc}) can be decomposed into components 
with definite $\varpi$-charge as 
\begin{eqnarray}
  \label{dcpqloc}
  &&\sum_{{n+m+r+s=K}\atop {m+s=L}}[
  C^a_{nm}(\kappa;\varpi+\kappa\lambda^a)
  C^a_{rs}(\nu;\varpi+\nu\lambda^a+\beta_{nm}^{a;\kappa})
  I^a_{nm}{}^{;a}_{;rs}(\kappa,\nu;\sigma,\sigma') \nonumber\\
  &&\hskip 2cm-C^a_{nm}(\kappa;\varpi+\kappa
  \lambda^a+\beta_{rs}^{a;\nu})C^a_{rs}(\nu;\varpi+\nu\lambda^a)
  I^a_{rs}{}^{;a}_{;nm}(\nu,\kappa;\sigma',\sigma)]=0 ~,\nonumber\\
  &&\sum_{{n+m+s=K}\atop {r+m+s=L}}[
  C^a_{nm}(\kappa;\varpi+\kappa\lambda^a)
  C^{\bar a}_{rs}(\nu;\varpi+\nu\lambda^{\bar a}+\beta_{nm}^{a;\kappa})
  I^{a,\kappa}_{nm}{}^{;{\bar a}\nu}_{;rs}(\sigma,\sigma')
  \nonumber\\
  &&\hskip 2cm-C^a_{nm}(\kappa;\varpi+
  \kappa\lambda^a+\beta_{rs}^{\bar a;\nu})
  C^{\bar a}_{rs}(\nu;\varpi+\nu\lambda^{\bar a})
  I^{\bar a,\nu}_{rs}{}^{;a}_{;nm}(\nu,\kappa;\sigma',\sigma)]=0 ~, 
\end{eqnarray}
where $I^a_{nm}{}^{;b}_{;rs}(\kappa,\nu;\sigma,\sigma')$ is defined by 
\begin{eqnarray}
  \label{Inmrs}
  I_{nm}^{a,\kappa}{}_{;rs}^{;b,\nu}(\sigma,\sigma')
  &=&V_\kappa^a(0,\sigma)({\cal Y}_a(0,\sigma))^n
  ({\cal Y}_{a\bar a}(0,\sigma))^m
  V_\nu^b(0,\sigma')({\cal Y}_b(0,\sigma'))^r
  ({\cal Y}_{b\bar b}(0,\sigma'))^s\nonumber\\
  &=&\psi_{nm}^{a+}(\kappa;\sigma)\psi_{rs}^{b+}(\nu;\sigma')\star
  \psi_{nm}^{a-}(\kappa;-\sigma)\psi_{rs}^{b-}(\nu;-\sigma')~,
\end{eqnarray}
and the sum should be taken over nonnegative integers $n,m,r,s$ for 
given $K,L$. 

For definiteness, we consider the case $0<\sigma'<\sigma<2\pi$. To 
reduce (\ref{dcpqloc}) further, let us choose an arbitrary point 
$\sigma''$ between $\sigma'$ and $\sigma$, and introduce operators by 
\begin{eqnarray}
  \label{yzs}
  && Y_1^\pm=\int_\sigma^{2\pi}dzV_a^\pm(\pm z)~, \quad
  Y_2^\pm=\int_{\sigma''}^\sigma dzV_a^\pm(\pm z)~, \nonumber\\
  && Z_1^\pm=\int_{\sigma'}^{\sigma''}dzV_{\bar a}^\pm(\pm z) ~, \quad
  Z_2^\pm=\int_0^{\sigma'}dzV_{\bar a}^\pm(\pm z) ~. 
\end{eqnarray}
These operators satisfy the following operator algebra 
\begin{eqnarray}
  \label{exchrel}
  &&Y_1^\pm Y_2^\pm=q^{\mp 2}Y_2^\pm Y_1^\pm ~, \quad
  Y_1^\pm Z_1^\pm=q^{\mp 1}Z_1^\pm Y_1^\pm ~, \quad
  Y_1^\pm Z_2^\pm=q^{\pm 1}Z_2^\pm Y_1^\pm ~, \nonumber\\
  &&\hskip 4.2cm Y_2^\pm  Z_1^\pm=q^{\pm 1}Z_1^\pm Y_2^\pm ~, \quad
  Y_2^\pm  Z_2^\pm=q^{\mp 1}Z_2^\pm Y_2^\pm ~, \\
  &&\hskip 8.4cm Z_1^\pm Z_2^\pm=q^{\mp 2}Z_2^\pm Z_1^\pm~.\nonumber
\end{eqnarray}
We also need the exchange algebra with the free field chiral vertex 
operators 
\begin{equation}
  \label{exchvyz}
\begin{array}{ll}
  Y_{1}^\pm V_\kappa^{a\pm}(\pm\sigma)=q^{\mp \kappa}
  V_\kappa^{a\pm}(\pm\sigma)Y_{1}^\pm ~, \quad 
  &Y_{2}^\pm V_\kappa^{a\pm}(\pm\sigma)=q^{\pm \kappa}
  V_\kappa^{a\pm}(\pm\sigma)Y_{2}^\pm ~, \\
  Y_{1,2}^\pm V_\kappa^{a\pm}(\pm\sigma')=q^{\mp\kappa}
  V_\kappa^{a\pm}(\pm\sigma')Y_{1,2}^\pm ~, \quad 
  & Z_{1,2}^\pm V_\kappa^{a\pm}(\pm \sigma)=
  V_\kappa^{a\pm}(\pm \sigma)Z_{1,2}^\pm ~,\\
  Y_{1,2}^\pm V_\nu^{\bar a\pm}(\pm \sigma)=
  V_\nu^{\bar a\pm}(\pm \sigma)Y_{1,2}^\pm ~,\quad  
  & Z_{1,2}^\pm V_\nu^{\bar a\pm}(\pm\sigma)=q^{\pm\nu}
  V_\nu^{\bar a\pm}(\pm\sigma)Z_{1,2}^\pm ~, \\
  Z_1^\pm V_\nu^{\bar a\pm}(\pm\sigma')=q^{\mp\nu}
  V_\nu^{\bar a\pm}(\pm\sigma')Z_1^\pm ~, \quad
  & Z_2^\pm V_\nu^{\bar a\pm}(\pm\sigma')=q^{\pm\nu}
  V_\nu^{\bar a\pm}(\pm\sigma')Z_2^\pm ~. 
\end{array}
\end{equation}
These relations are used to move the vertex operators from the right 
to the left of $Y_k^\pm$ and $Z_k^\pm$ or vice versa 
in operator products.

We single out the contributions from the operators (\ref{yzs}) to the 
screening charges and neglect all others as 
\begin{equation}
  \label{scrchYZ}
  \begin{array}{ll}
  A_a(\sigma)\sim  f^+(\varpi^a+1)~,\quad 
  &A_{a\bar a}(\sigma)\sim q^{\varpi^{\bar a}}f^+(\varpi^a
  +\varpi^{\bar a}+1)g^+(0) ~,\\
  A_a(\sigma')\sim  q^{-\varpi^a-1}f^+(0)~,\quad 
  & A_{a\bar a}(\sigma')\sim q^{-\varpi^a-1}f^+(0)g^+(0) ~, \\
  A_{\bar a}(\sigma)\sim q^{\varpi^{\bar a}+1}g^+(0)~, \quad
  & A_{\bar aa}(\sigma)\sim q^{\varpi^{\bar a}+1}g^+(0)f^+(0)~,\\
  A_{\bar a}(\sigma')\sim g^+(\varpi^{\bar a}+1)~, \quad
  & A_{\bar aa}(\sigma')\sim q^{-\varpi^a}g^+(\varpi^a+\varpi^{\bar a}+1)
  f^+(0)~, \\
  B_a(-\sigma)\sim f^-(\varpi^a+1) ~, \quad 
  & B_{a\bar a}(-\sigma)\sim q^{-\varpi^{\bar a}}
  f^-(\varpi^a+\varpi^{\bar a}+1)g^-(0) ~, \\
  B_a(-\sigma')\sim q^{\varpi^a+1}g^-(\varpi^{\bar a}+1) ~, \quad 
  &B_{a\bar a}(-\sigma')\sim q^{\varpi^a+1}
  f^-(0)g^-(0) ~, \\
  B_{\bar a}(-\sigma)\sim q^{-\varpi^{\bar a}-1}g^-(0) ~, \quad
  & B_{\bar aa}(-\sigma)\sim q^{-\varpi^{\bar a}-1}g^-(0)f^-(0) ~, \\
  B_{\bar a}(-\sigma')\sim g^-(\varpi^{\bar a}+1) ~, \quad
  & B_{\bar aa}(-\sigma')\sim q^{\varpi^a}g^-(\varpi^a+\varpi^{\bar a}+1)
  f^-(0) ~, 
\end{array}
\end{equation}
where $f^\pm$ and $g^\pm$ are defined by 
\begin{eqnarray}
  \label{fg}
  f^\pm(\xi)=q^{\mp\xi}Y^\pm_1+q^{\pm\xi}Y^\pm_2 ~, \qquad
  g^\pm(\xi)=q^{\mp\xi}Z^\pm_1+q^{\pm\xi}Z^\pm_2 ~.
\end{eqnarray}
Note the factorized forms for $A_{ab}$ and $B_{ab}$. The vector forms
of the screening charges ${\cal Y}_a$ and ${\cal Y}_{ab}$ are more 
convenient for the locality analysis. We can easily derive their 
truncated expressions from (\ref{scrchYZ}) as
\begin{eqnarray}
  \label{tranys}
  \begin{array}{ll}
  {\cal Y}_a(0,\sigma)\sim f(\varpi^a+1)~, &
  {\cal Y}_{a\bar a}(0,\sigma)\sim f(\varpi^a+\varpi^{\bar a}+1)g(0)~,\\
  {\cal Y}_{\bar a}(0,\sigma)\sim g(0)~, &
  {\cal Y}_{\bar aa}(0,\sigma)\sim g(0)f(0)~, \\
  {\cal Y}_{a}(0,\sigma')\sim f(0) ~, &
  {\cal Y}_{a\bar a}(0,\sigma')\sim f(0)g(0)~, \\
  {\cal Y}_{\bar a}(0,\sigma')\sim g(\varpi^{\bar a}+1)~, &
  {\cal Y}_{\bar aa}(0,\sigma')\sim g(\varpi^a+\varpi^{\bar a}+1)f(0)~, 
  \end{array}
\end{eqnarray}
where $f$ and $g$ are defined by
\begin{eqnarray}
  \label{vfg}
  f(\xi)=f^+(\xi)\star f^-(\xi)~, \qquad 
  g(\xi)=g^+(\xi)\star g^-(\xi)~.
\end{eqnarray}

The operator algebra (\ref{exchrel}) possesses a realization in terms 
of the zero-mode operators $Q$ and $P$. Let us introduce operators by 
\begin{eqnarray}
  \label{taua}
  \Lambda_a={\rm e}^{\gamma\eta\alpha^a\cdot Q}~,\quad 
  y^\pm=\frac{1}{3}(2\varpi^a+\varpi^{\bar a})+y_0^\pm~, \quad
  z^\pm=\frac{1}{3}(\varpi^a+2\varpi^{\bar a})+z_0^\pm~,
\end{eqnarray}
where $y_0^\pm$ and $z_0^\pm$ are arbitrary constants. Then the operator
algebra  (\ref{exchrel}) is realized by the operators\footnote{We use 
the same notation for the quantum mechanical realization.} defined by 
\begin{eqnarray}
  \label{qmr}
  Y_1^\pm= \Lambda_a~, \quad 
  Y_2^\pm=-q^{\mp2y^\pm}\Lambda_a~, \quad
  Z_1^\pm= q^{\mp\varpi^{\bar a}}\Lambda_{\bar a}~, \quad
  Z_2^\pm=-q^{\mp\varpi^{\bar a}}q^{\mp2z^\pm}\Lambda_{\bar a}~.
\end{eqnarray}
This can be seen by noting that the operators given by (\ref{taua}) 
satisfy
\begin{eqnarray}
  \label{opalgyz}
  \begin{array}{lll}
  \Lambda_ay^\pm=(y^\pm+1)\Lambda_a ~, 
  &&\Lambda_{\bar a}y^\pm=y^\pm\Lambda_{\bar a} ~,\\
  \Lambda_az^\pm=z^\pm\Lambda_a ~, 
  &&\Lambda_{\bar a}z^\pm=(z^\pm+1)\Lambda_{\bar a}~.
  \end{array}
\end{eqnarray}
After the substitution of (\ref{qmr}) into (\ref{fg}) and (\ref{vfg}), we 
arrive at interesting expressions for $f$ and $g$ as 
\begin{eqnarray}
  \label{qmrfg}
  f(\xi)=[\xi-y^+][\xi-y^-]\tilde\Lambda_a~, \qquad
  g(\xi)=[\xi-z^+][\xi-z^-]\tilde\Lambda_{\bar a}~,
\end{eqnarray}
where we have introduced $q$-numbers by
\begin{eqnarray}
  \label{qn}
  [x]=\frac{q^x-q^{-x}}{q-q^{-1}}~,
\end{eqnarray}
and $\tilde\Lambda$ is defined by 
\begin{eqnarray}
  \label{tlam}
  \tilde\Lambda_a=-q^{-y_0^++y_0^-}(q-q^{-1})^2\Lambda_a ~, \qquad
  \tilde\Lambda_{\bar a}
  =-q^{-z_0^++z_0^-}(q-q^{-1})^2\Lambda_{\bar a} ~.
\end{eqnarray}

Due to the independence of the chiral vertex operators $V_a^\pm$ at different 
points, the locality constraints (\ref{dcpqloc}) should hold true even if we 
replace the screening charges by their truncated forms given by 
(\ref{tranys}). We apply this idea of truncation to find the expansion 
coefficients. 
As we shall see shortly, any coefficients $C_{nm}^a$ can be expressed in 
terms of $C^a_{n0}$ by solving the second conditions of (\ref{dcpqloc}), 
whereas they are obtained from the first of (\ref{dcpqloc}) for $L=0$. 
It is essentially equivalent to the locality constraints for the 
Liouville theory and has been investigated in ref. \cite{fit96}. 

To illustrate our 
method we solve here the Liouville case in the present notation. 
From (\ref{tranys}), the first constraint of (\ref{dcpqloc}) for $L=0$ 
can be cast into the following form
\begin{eqnarray}
  \label{flocL=0}
  &&\sum_{n+r=K}\{C^a_{n0}(\kappa;\varpi-\nu\lambda^a)
  C^a_{r0}(\nu;\varpi+n\alpha^a)(f(\varpi^a-\nu+1))^n (f(0))^r \nonumber\\
  &&\hskip 2cm -C^a_{n0}(\kappa;\varpi+r\alpha^a)
  C^a_{r0}(\nu;\varpi-\kappa\lambda^a)(f(\kappa))^r
  (f(\varpi^a+1))^n\}=0 ~,
\end{eqnarray}
where we have removed the vertex operators $V_\kappa^a$ and $V_\nu^a$ 
after moving them to the left of the expressions by using (\ref{exchvyz}).
At this point we go over to the quantum mechanical realization (\ref{qmr}) 
and use (\ref{qmrfg}). We then remove all the $\tilde\Lambda$ after 
moving them to the right of the expression. This leads to an equivalent 
form 
\begin{eqnarray}
  \label{flocL2=0}
  &&\sum_{n+r=K}\{C^a_{n0}(\kappa;\varpi-\nu\lambda^a)
  C^a_{r0}(\nu;\varpi+n\alpha^a)\nonumber\\
  &&\hskip 2cm \times[\varpi^a-\nu+1-y^+]_n
  [\varpi^a-\nu+1-y^-]_n[y^++n]_r[y^-+n]_r \nonumber\\
  &&\hskip 1cm -C^a_{n0}(\kappa;\varpi+r\alpha^a)
  C^a_{r0}(\nu;\varpi-\kappa\lambda^a)\nonumber\\
  &&\hskip 2cm \times[\varpi^a+r+1-y^+]_n
  [\varpi^a+r+1-y^-]_n[y^+-\kappa]_r[y^--\kappa]_r\}=0 ~,
\end{eqnarray}
where $[x]_n\equiv[x][x+1]\cdots[x+n-1]$ with $[x]_0=1$. 
Since $y^\pm$ are arbitrary, we may freely choose them 
to find $C_{n0}^a$. For $y^+=1-K$ and $y^-=\kappa$ (\ref{flocL2=0}) 
reduces to 
\begin{eqnarray}
  \label{step1}
  &&C_{K0}^a(\kappa;\varpi-\nu\lambda^a)[\varpi^a-\nu+K]_K
  [\varpi^a-\kappa-\nu+1]_K \nonumber\\
  &&\hskip 2.0cm =C_{K0}^a(\kappa;\varpi)[\varpi^a+K]_K
  [\varpi^a-\kappa+1]_K~.
\end{eqnarray}
Since $C^a_{n0}$ does not depend on $\varpi^{\bar a}$ as we shall 
show below, this completely fix the $\varpi$ depedence as
\begin{eqnarray}
  \label{cn01}
  C^a_{n0}(\kappa;\varpi)=\frac{f_n(\kappa)}{[\varpi^a+n]_n
  [\varpi^a-\kappa+1]_n}~,
\end{eqnarray}
where $f_n(\kappa)$ is a function of $\kappa$. To determine $f_n$ 
we next assume $y^+=\varpi^a-\nu+1$ and $y^-=\kappa$. Then 
(\ref{flocL2=0}) and (\ref{cn01}) give $f_K(\kappa)[\nu]_K
=f_K(\nu)[\kappa]_K$. Hence we may put 
\begin{eqnarray}
  \label{fn}
  f_n(\kappa)=c_n[\kappa]_n~,
\end{eqnarray}
where $c_n$ is independent of $\kappa$. It satisfies the recurrence 
relation $[n]c_n=c_1c_{n-1}$. This can be obtained from (\ref{flocL2=0})
combined with (\ref{cn01}) and (\ref{fn}) for $y^+=1-K$ and $y^-=\kappa-1$.
Since $c_0=1$ by definition, we find 
\begin{eqnarray}
  \label{cn}
  c_n=\frac{c_1^n}{[n]!}~, \qquad (n=0,1,\cdots)
\end{eqnarray}
where $[n]!\equiv[1][2]\cdots[n]$ is the $q$-factorial and $c_1$ is not 
fixed by locality constraints. It is determined by the 
requirement of the field equation as 
\begin{eqnarray}
  \label{c1}
  c_1=\frac{\eta}{8\pi g\sin2\pi g}~.
\end{eqnarray}
We will show this in the next section. Combining all these results, 
we obtain $C^a_{n0}$ as 
\begin{eqnarray}
  \label{can0}
  C^a_{n0}(\kappa;\varpi)=\Biggl(\frac{\eta}{8\pi g\sin2\pi g}\Biggr)^n
  \frac{[\kappa]_n}{[n]![\varpi^a+n]_n
  [\varpi^a-\kappa+1]_n}~.
\end{eqnarray}

We now turn to the second condition of (\ref{dcpqloc}). It can be 
dealt with by the similar method as above. We substitute (\ref{tranys}) 
into the condition and then moving the free vertex operators $V_\kappa^a$ 
and $V_\nu^{\bar a}$ to the left of the expression. After the manipulation 
we can safely remove them and the quantum mechanical realization 
(\ref{qmrfg}) can be applied. We gather all the $\tilde\Lambda$ to the 
right of the expression by noting (\ref{opalgyz}) and then remove them from 
the equation. The truncated locality condition can thus be cast into the 
following form
\begin{eqnarray}
  \label{2ndlocc}
  &&\sum_{{n+m+s=K}\atop {r+m+s=L}}\{
  C^a_{nm}(\kappa;\varpi-\nu\lambda^{\bar a})
  C^{\bar a}_{rs}(\nu;\varpi+(n+m)\alpha^a+m\alpha^{\bar a})
   \nonumber\\
  &&\hskip 2.5cm \times
  [\varpi^a+1-y^+]_n[\varpi^a+1-y^-]_n \nonumber\\
  &&\hskip 2.5cm \times[\varpi^a+\varpi^{\bar a}-\nu+1-y^+]_m
  [\varpi^a+\varpi^{\bar a}-\nu+1-y^-]_m[z^+-\nu]_m[z^--\nu]_m
   \nonumber\\
  &&\hskip 2.5cm \times 
  [\varpi^{\bar a}-n+1-z^+]_r[\varpi^{\bar a}-n+1-z^-]_r \nonumber\\
  &&\hskip 2.5cm \times [y^++n+m]_s[y^-+n+m]_s
  \nonumber\\
  &&\hskip 2.5cm \times 
  [\varpi^a+\varpi^{\bar a}+n+m+1-z^+]_s[\varpi^a+\varpi^{\bar a}+n+m+1-z^-]_s
  \nonumber\\
  &&\hskip 1.4cm-C^a_{nm}(\kappa;\varpi+s\alpha^a+(r+s)\alpha^{\bar a})
  C^{\bar a}_{rs}(\nu;\varpi-\kappa\lambda^a)\nonumber\\
  &&\hskip 2.5cm \times [\varpi^a-r+1-y^+]_n[\varpi^a-r+1-y^-]_n \nonumber\\
  &&\hskip 2.5cm \times[\varpi^a+\varpi^{\bar a}+r+s+1-y^+]_m
  [\varpi^a+\varpi^{\bar a}+r+s+1-y^-]_m \nonumber\\
  &&\hskip 2.5cm \times[z^++r+s]_m[z^-+r+s]_m \nonumber\\
  &&\hskip 2.5cm \times[\varpi^{\bar a}+1-z^+]_r[\varpi^{\bar a}+1-z^-]_r 
  [y^+-\kappa]_s[y^--\kappa]_s
  \nonumber\\
  &&\hskip 2.5cm \times[\varpi^a+\varpi^{\bar a}-\kappa+1-z^+]_s
  [\varpi^a+\varpi^{\bar a}-\kappa+1-z^-]_s\}  =0 ~.   
\end{eqnarray}
We first confirm that the $C^a_{n0}$ is independent of $\varpi^{\bar a}$ as 
announced before. For $L=0$,  (\ref{2ndlocc}) reduces to 
\begin{eqnarray}
  \label{wabindep}
  C^a_{K0}(\kappa;\varpi-\nu\lambda^{\bar a})=C^a_{K0}(\kappa;\varpi)~,
\end{eqnarray}
which implies that the $C^a_{n0}$ depends only on $\varpi^a$. 

We now consider the case $K\ge L$ and put $n=K-L$, $m=L$. For the choice of 
the arbitrary parameters
\begin{eqnarray}
  \label{ypmzpm}
  y^+=\kappa~, \quad y^-=1-n-m~, \quad z^+=\nu~, \quad 
  z^-=\varpi^{\bar a}+1~,
\end{eqnarray}
we obtain from (\ref{2ndlocc})
\begin{eqnarray}
  \label{cnm-cl0}
  C^a_{nm}(\kappa;\varpi)&=&(-1)^m\frac{[n+m]_m[\varpi^a+2n+m]_m
    [\varpi^a-\kappa+n+1]_m[\varpi^{\bar a}-\nu-n-m+1]_m}{%
    [\nu]_m[\varpi^a+\varpi^{\bar a}+n+m]_m
    [\varpi^a+\varpi^{\bar a}-\kappa+1]_m[\varpi^{\bar a}+1]_m}\nonumber\\
  &&\times C^a_{n+m\:0}(\kappa;\varpi-\nu\lambda^{\bar a})
  C^{\bar a}_{m\:0}(\nu;\varpi+(n+m)\alpha^a) ~.
\end{eqnarray}
We see that the coefficients $C^a_{nm}$ with $m\ne0$, which are 
characteristic of $A_2$-Toda theory, are related to those of $A_1$ 
case, i.e., the Liouville theory. This combined with (\ref{can0}) 
leads to the full expression for the expansion coefficients 
\begin{eqnarray}
  \label{canm}
  C^a_{nm}(\kappa;\varpi)&=&(-1)^m
  \Biggl(\frac{\eta}{8\pi g\sin2\pi g}\Biggr)^{n+2m}
  \frac{[\kappa]_{n+m}}{[n]![m]!}
  \frac{1}{[\varpi^a+n+m]_n[\varpi^a-\kappa+1]_n}\nonumber\\
  &&\times \frac{1}{
    [\varpi^a+\varpi^{\bar a}+n+m]_m
    [\varpi^a+\varpi^{\bar a}-\kappa+1]_m
    [\varpi^{\bar a}-n]_m[\varpi^{\bar a}+1]_m}~.
\end{eqnarray}
Except for the apparent difference due to the slight change of 
convention, this coincides to the the conjecture given in ref. 
\cite{fit98}. It is easy to verify that (\ref{canm}) satisfies 
the conditions (\ref{Cscon}). 

We can interpret (\ref{cnm-cl0}) as quantum deformation of the 
classical Toda exponential function. This can be seen by 
taking the classical limit, which is defined by $g\rightarrow0$ 
with $\psi_\pm$ kept fixed. Since $[c]\rightarrow c$ and 
$2\pi g\:[\beta\cdot\varpi+c]\rightarrow -i\:{\rm sinh}
\displaystyle{\frac{\gamma}{4}}\beta\cdot P$ for any 
constant $c$, we can easily compute the classical limit of 
(\ref{cnm-cl0}) as
\begin{eqnarray}
  \label{climcanm}
  C^a_{nm}(\kappa;\varpi)&\rightarrow&{-\kappa\choose n+m}
  {n+m\choose n}(C_{\alpha^a})^{2n}(C_{\alpha^a+\alpha^{\bar a}}
  C_{\alpha^{\bar a}})^{2m} ~,
\end{eqnarray}
where $\displaystyle{\nu\choose n}$ is the ordinary binomial 
coefficient and $C_\beta$ is given by (\ref{ce}). 
This coincides with the coefficients appearing in the expansion 
of the classical Toda exponential 
${\rm e}^{\kappa\lambda^a\cdot\varphi}$ obtained from 
(\ref{a2csol}), where the coefficients $C$'s are contained 
in the screening charges. 

If $\kappa=-j$ with some nonnegative integer $j$, $[-j]_{n+m}=0$
for $n+m>j$. Hence, ${\rm e}^{-\eta j\lambda^a\cdot\varphi}$ 
reduces to a finite polynomial in the screening charges as 
is expected from the classical solution (\ref{a2csol}) \cite{fl,bg,bg89}. 
It contains $\frac{1}{2}(j+1)(j+2)$ terms corresponding to 
the dimension of the completely symmetric tensor product 
of $j$ defining representations. 

So far our main concern is the construction of the Toda exponential
operators associtated with the fundamental weights. Arbitrary 
exponential operators can be constructed from them as composite 
operators. Let $\beta$ be an arbitrary vector in the $A_2$ root space, 
then it can be expressed as a linear combination of the fundamental 
weights as
\begin{eqnarray}
  \label{bll}
  \beta=\beta^a+\beta^{\bar a} \qquad 
  {\rm with}\quad \beta^a\equiv\lambda^a\alpha^a\cdot\beta
  ~, \quad \beta^{\bar a}\equiv\lambda^{\bar a}\alpha^{\bar a}\cdot\beta ~.
\end{eqnarray}
Hence we may define
\begin{eqnarray}
  \label{expbphi}
  {\rm e}^{\eta\beta\cdot\varphi(x)}
  &=&\lim_{x'\to x}|(1-{\rm e}^{-i(x^+-x'{}^+)})
  (1-{\rm e}^{-i(x^--x'{}^-)})|^{\Delta_\beta-\Delta_{\beta^a}
    -\Delta_{\beta^{\bar a}}}
  {\rm e}^{\eta\beta^a\cdot\varphi(x)}
  {\rm e}^{\eta\beta^{\bar a}\cdot\varphi(x')} \nonumber\\
  &=&\sum_{n,m}\sum_{r,s}C^a_{nm}(\alpha^a\cdot\beta;\varpi
  +\beta^a)C^{\bar a}_{rs}(\alpha^{\bar a}\cdot\beta;
  \varpi+\beta+(n+m)\alpha^a+m\alpha^{\bar a})\nonumber\\
  &&\hskip 1.5cm\times({\cal Y}_a(x))^n({\cal Y}_{a\bar a}(x))^m
  :{\rm e}^{\eta\beta\cdot\psi(x)}:({\cal Y}_{\bar a}(x))^r
  ({\cal Y}_{\bar aa}(x))^s ~,
\end{eqnarray}
where $\Delta$'s are the conformal weights of the exponential operators 
and are given by (\ref{cdim}). In particular, we need 
such operators associated with the simple roots to establish 
the operatorial field equations. From the general formula 
(\ref{expbphi}) we obtain for $\beta=\alpha^a=2\lambda^a-\lambda^{\bar a}$
\begin{eqnarray}
  \label{ealp}
  {\rm e}^{\eta\alpha^a\cdot\varphi(x)}&=&\sum_{n,m}
  \sum_{r+s\le1}
  C^a_{nm}(2;\varpi+2\lambda^a)C^{\bar a}_{rs}(-1;\varpi+(n+m+1)
  \alpha^a+m\alpha^{\bar a})\nonumber\\
  &&\hskip 1.5cm\times({\cal Y}_a(x))^n({\cal Y}_{a\bar a}(x))^m
  V_a(x)({\cal Y}_{\bar a}(x))^r
  ({\cal Y}_{\bar aa}(x))^s ~, 
\end{eqnarray}
where $V_a(x)=V_a^+(x^+)\star V_a^-(x^-)$. This will be used in the next 
section.

\section{Field Equations}
\label{sec:feq}
\setcounter{equation}{0}

From the Toda exponential operator (\ref{teo}) one can define the local 
Toda field operators as the derivative with respect to $\kappa$ at 
$\kappa=0$. It is explicitly given by
\begin{eqnarray}
  \label{ltf}
  \eta\lambda^a\cdot\varphi=\eta\lambda^a\cdot\psi-\frac{1}{\sin^22\pi g}
  \sum_{n,m}\Biggl(\frac{\mu^2}{4}\Biggr)^{n+2m}D^a_{nm}(\varpi)
  {\cal Y}_a^n{\cal Y}_{a\bar a}^m~,
\end{eqnarray}
where $D^a_{nm}$ is defined by 
\begin{eqnarray}
  \label{danm}
  D^a_{nm}(\varpi)&=&-\sin^22\pi g\left.\frac{d}{d\kappa}
    C^a_{nm}(\kappa;\varpi)
  \right|_{\kappa=0}
  \nonumber\\
  &=&(-1)^{m-1}\frac{c_1^{n+2m-1}}{4}\frac{[n+m-1]!}{[n]![m]!}
  \frac{1}{[\varpi^a+n+m]_n[\varpi^a+1]_n}\nonumber\\
  &&\times \frac{1}{[\varpi^a+\varpi^{\bar a}+n+m]_m[\varpi^a+\varpi^{\bar a}
    +1]_m[\varpi^{\bar a}-n]_m[\varpi^{\bar a}+1]_m} ~.
\end{eqnarray}
We show that the Toda fields thus defined satisfy the operator field equations 
\begin{eqnarray}
  \label{fieldeq}
  \partial_\mu\partial^\mu\varphi
  +\mu^2\sum_{a=1,2}\alpha^a{\rm e}^{\eta\alpha^a\cdot\varphi}=0~,
\end{eqnarray}
where the exponential operators associated with the simple roots are 
defined by (\ref{ealp}). By decomposing (\ref{fieldeq}) into sectors 
with definite $\varpi$-charges, we can cast (\ref{fieldeq}) into an 
equivalent from 
\begin{eqnarray}
  \label{fequiv}
  &&\frac{1}{4\sin^22\pi g}\partial_+\partial_-({\cal Y}_a^{n+1}
  {\cal Y}_{a\bar a}^m) \nonumber\\
  &&\hskip 1cm=-[n+1][n+m+1]
    [\varpi^a+n+1][\varpi^a+n+m+1] \nonumber\\
    &&\hskip 1.8cm\times\frac{[\varpi^a+\varpi^{\bar a}+n+m+1]
    [\varpi^{\bar a}-n-1]}{[\varpi^a+\varpi^{\bar a}+n+2m+1]
    [\varpi^{\bar a}-n+m-1]}
  {\cal Y}_a^n{\cal Y}_{a\bar a}^mV_a\nonumber\\
  &&\hskip 1cm-[m][n+m+1][\varpi^a+\varpi^{\bar a}+n+m+1]
  [\varpi^a+\varpi^{\bar a}+m]\nonumber\\
  &&\hskip 1.8cm
  \times\frac{[\varpi^a+n+m+1][\varpi^{\bar a}+m]}{[\varpi^a+2n+m+2]
    [\varpi^{\bar a}-n+m-1]}{\cal Y}_a^{n+1}{\cal Y}_{a\bar a}^{m-1}
  V_a{\cal Y}_{\bar a}\nonumber\\
  &&\hskip 1cm+[n+1][m][\varpi^{\bar a}-n-1][\varpi^{\bar a}+m]\nonumber\\
  &&\hskip 1.8cm\times
  \frac{[\varpi^a+n+1][\varpi^a+\varpi^{\bar a}+m]}{[\varpi^a+2n+m+2]
    [\varpi^a+\varpi^{\bar a}+n+2m+1]} {\cal Y}_a^n{\cal Y}_{a\bar a}^{m-1}
  V_a{\cal Y}_{\bar aa}~,
\end{eqnarray}
where use has been made of the explicit forms (\ref{canm}) and (\ref{danm}). 

To show (\ref{fequiv}) we first note the relations 
\begin{eqnarray}
  \label{dav}
  \partial_+A_a=2i\sin2\pi g[\varpi^a+1]V_a^+~, \qquad
  \partial_+A_{a\bar a}=2i\sin2\pi g[\varpi^a+\varpi^{\bar a}+1]V_a^+
  A_{\bar a}~.
\end{eqnarray}
Since $V_a^+$ commutes with $A_a$, we find   
\begin{eqnarray}
  \label{davn}
  \partial_+A_a^{n+1}=2i\sin2\pi g[n+1][\varpi^a+n+1]A_a^nV_a^+~.
\end{eqnarray}
In deriving this, use has been made of the relation
\begin{eqnarray}
  \label{sumqn}
  \sum_{k=0}^n[x+2k+1]=[n+1][x+n+1]~.
\end{eqnarray}
Similarly, from the commutativity of $V_a^+A_{\bar a}$ and $A_{a\bar a}$, 
we obtain 
\begin{eqnarray}
  \label{daabam}
  \partial_+A_{a\bar a}^m=2i\sin2\pi g[m][\varpi^a+\varpi^{\bar a}+m]
  A_{a\bar a}^{m-1}V_a^+A_{\bar a}~.
\end{eqnarray}
This combined with (\ref{davn}) yields 
\begin{eqnarray}
  \label{daanaabam}
  \frac{1}{2i\:\sin2\pi g}\partial_+(A_a^{n+1}A_{a\bar a}^m)
  &=&\frac{[n+1][\varpi^a+n+1][\varpi^{\bar a}-n-1]}{[\varpi^{\bar a}-n+m-1]}
  A_a^nA_{a\bar a}^mV_a^+\nonumber\\
  &&\hskip .5cm+\frac{[m][\varpi^a+\varpi^{\bar a}+m]
    [\varpi^{\bar a}+m]}{[\varpi^{\bar a}-n+m-1]}
  A_a^{n+1}A_{a\bar a}^{m-1}V_a^+A_{\bar a}~,
\end{eqnarray}
where use has been made of the relation
\begin{eqnarray}
  \label{vaaabam}
  V_a^+A_{a\bar a}^m=\frac{[\varpi^{\bar a}-1]}{[\varpi^{\bar a}+m+1]}
  A_{a\bar a}^m+\frac{[m]}{[\varpi^{\bar a}+m+1]}A_aA_{a\bar a}^{m-1}
  V_a^+A_{\bar a} ~.
\end{eqnarray}
This can be shown by mathematical induction. It is now straightforwards to 
verify (\ref{fequiv}) from (\ref{daanaabam}) and the analogous relation 
for the right-chiral sector obtained by the replacements $V_a^+\rightarrow 
V_a^-$, $A\rightarrow B$ and $\partial_+\rightarrow\partial_-$.

We now turn our attention to the lhs of (\ref{fequiv}) and evaluate it as  
\begin{eqnarray}
  \label{dynym}
  \frac{1}{4\sin^22\pi g}\partial_+\partial_-({\cal Y}_a^{n+1}
  {\cal Y}_{a\bar a}^m)&=&\alpha(\varpi)^2{\cal Y}_a^n{\cal Y}_{a\bar a}^m
  V_a+\beta(\varpi)^2{\cal Y}^{n+1}{\cal Y}_{a\bar a}^{m-1}
  V_a{\cal Y}_{\bar a} \nonumber\\
  &&+\alpha(\varpi)\beta(\varpi)
  {\cal Y}_a^n{\cal Y}_{a\bar a}^{m-1}
  (A_{a\bar a}V_a^+\star V_a^-B_aB_{\bar a}
  +V_aA_aA_{\bar a}\star B_{a\bar a}V_a^-)~, \nonumber\\
\end{eqnarray}
where $\alpha$ and $\beta$ are given by
\begin{eqnarray}
  \label{alpbet}
  \alpha(\varpi)\equiv\frac{[n+1][\varpi^a+n+1]
    [\varpi^{\bar a}-n-1]}{[\varpi^{\bar a}-n+m-1]} ~,\quad
  \beta(\varpi)\equiv\frac{[m][\varpi^a+\varpi^{\bar a}+m]
    [\varpi^{\bar a}+m]}{[\varpi^{\bar a}-n+m-1]}~.
\end{eqnarray}
To simplify the crossed terms on the rhs of (\ref{dynym}) further, 
we note the relation
\begin{eqnarray}
  \label{va}
  V_a^+A_{\bar aa}=\gamma(\varpi)A_{a\bar a}V^+_a+\delta(\varpi)
  A_aA_{\bar a}V^+_a~,
\end{eqnarray}
where $\gamma$ and $\delta$ are defined by
\begin{eqnarray}
  \label{adbd}
  \gamma(\varpi)\equiv-\frac{[\varpi^a+3]}{[\varpi^{\bar a}]}~, \qquad
    \delta(\varpi)\equiv\frac{[\varpi^a
      +\varpi^{\bar a}+3]}{[\varpi^{\bar a}]}~.
\end{eqnarray}
Eq. (\ref{va}) and the analogous relation for the right-chiral 
operators give 
\begin{eqnarray}
  \label{vy}
  V_a{\cal Y}_{\bar aa}&=&\gamma(\varpi)^2{\cal Y}_{a\bar a}
  V_a+\delta(\varpi)^2{\cal Y}_aV_a{\cal Y}_{\bar a} \nonumber\\
  &&+\gamma(\varpi)\delta(\varpi)
  (A_{a\bar a}V_a^+\star V_a^-B_aB_{\bar a}
  +V_aA_aA_{\bar a}\star B_{a\bar a}V_a^-) ~.
\end{eqnarray}
From (\ref{dynym}) and (\ref{vy}) we arrive at the following 
expression
\begin{eqnarray}
  \label{dynym2}
  \frac{1}{4\sin^22\pi g}\partial_+\partial_-({\cal Y}_a^{n+1}
  {\cal Y}_{a\bar a}^m)&=&-\alpha(\varpi)\Biggl(
  \alpha(\varpi)-\beta(\varpi)
  \frac{\delta(\tilde\varpi)}{\gamma(\tilde\varpi)}\Biggr)
  {\cal Y}_a^n{\cal Y}_{a\bar a}^mV_a \nonumber\\
  &&-\beta(\varpi)\Biggl(\beta(\varpi)-\alpha(\varpi)
  \frac{\gamma(\tilde\varpi)}{\delta(\tilde\varpi)}\Biggr)
  {\cal Y}_a^{n+1}{\cal Y}_{a\bar a}^{m-1}
  V_a{\cal Y}_{\bar a}\nonumber\\
  &&-\frac{\alpha(\varpi)\beta(\varpi)}{\gamma(\tilde\varpi)
  \delta(\tilde\varpi)}
  {\cal Y}_a^n{\cal Y}_{a\bar a}^{m-1}V_a{\cal Y}_{\bar aa}
\end{eqnarray}
with $\tilde\varpi=\varpi+(n+m-1)\alpha^a+(m-1)\alpha^{\bar a}$.
Using the explicit forms (\ref{alpbet}) and (\ref{adbd}), one 
can show that (\ref{dynym2}) is nothing but eq. (\ref{fequiv}). 
This completes the proof of the operator field equations 
(\ref{fieldeq}). 

\section{Quantum Exchange Algebra and Locality}
\label{sec:qealgandloc}
\setcounter{equation}{0}

We have shown that the requirement of locality for (\ref{teo}) 
determines the coefficients $C^a_{nm}$. In fact, the truncated 
screening charges (\ref{scrchYZ}) suffice to solve the locality 
conditions (\ref{dcpqloc}). Since the operator algebra (\ref{oep}) 
of the full screening charges gives stronger restrictions on the 
coefficients $C^a_{nm}$ than that 
of the truncated ones, it might occur that (\ref{dcpqloc}) would 
not be satisfied for any choice of the coefficients. So, we should 
confirm the locality with the coefficients (\ref{canm}). One way 
to achieve this goal is to show that the ${\cal R}$-matrix 
satisfies (\ref{locR2}). As we have argued in sect. \ref{sec:qa2toda}, 
they are equivalent to the locality. It is, 
however, a rather involved task to obtain the explicit forms of 
the ${\cal R}$-matrices from the exchange algebra (\ref{oep}). 
Fortunately, there are cases where they can be found or 
be systematically constructed by the method developed in 
sect. \ref{sec:qteo}. 

We first consider the exchange algebra (\ref{excalg}) for 
$b=a$ in the left-chiral sector. We further assume $x>x'$ 
so that the decompositions of the screening charges introduced 
in sect. \ref{sec:qteo} can be used. We substitute the 
screening charges by the truncated ones given by 
(\ref{scrchYZ}) and then apply the algebraic manipulations 
leading to (\ref{flocL2=0}). We thus obtain 
\begin{eqnarray}
  \label{rmataa}
  &&\sum_{r+r'=n+n'\atop s+s'=m+m'}(-1)^{r'+s}
  q^{-(r'+s')(\varpi^a+r'+s')+s(\varpi^{\bar a}-r'+s')
    +\frac{2}{3}\kappa\nu+(r+s)\nu-s(r+s')}
  \nonumber\\
  &&\hskip 1.5cm\times 
  {\cal R}_{nm}^{rs}{}_{;n'm'}^{;r's'}
  ({\scriptstyle {a\atop 
      \kappa}}|{\scriptstyle {a\atop \nu}};\varpi\bigr)
  [y]_{r'+s'}[\varpi^a+\nu+1+r'-y]_r
  [\varpi^a+\varpi^{\bar a}+\nu+1+s'-y]_s\nonumber\\
  &&\hskip 2.5cm=(-1)^{n'+m}q^{-(n'+m')(\varpi^a+\kappa
    +2n+m+n'+m')+m(\varpi^{\bar a}-n)-m(n'+m')}\nonumber\\
  &&\hskip 3.5cm\times[\varpi^a+1-y]_n[\varpi^a+\varpi^{\bar a}+1-y]_m
  [\kappa+n+m+y]_{n'+m'}~,
\end{eqnarray}
where $y\equiv y^+$ is an arbitrary parameter. Since this gives only 
$n+m+n'+m'+1$ relations for $(n+n'+1)(m+m'+1)$ elements of the 
${\cal R}$-matrix for fixed $n,m,n',m'$, we can not determine all the 
elements for $(n+n')(m+m')\ne0$. This is due to the 
oversimplification by the use of the truncated screening charges, and 
merely implies that the truncation of the screening charges does not 
work well here.\footnote{In arriving  
at (\ref{rmataa}) we divide out the common factor $[z^+]_{n+n'}$
originating from the truncated screening charge $(g^+(0))^{n+n'}$.}
One can, however, find the matrix elements for $m=m'=0$ 
or $n=n'=0$ from (\ref{rmataa}). The former corresponds essentially 
to the Liouville case and has already been investigated in ref. 
\cite{gs93}. On the other hand the latter case is specific to 
$A_2$-Toda theory. Interestingly, these two cases can be described by
$A_{nm}^{rs}$ defined by the equation
\begin{eqnarray}
  \label{anmrs}
  \sum_{r+s=n+m}A_{nm}^{rs}(\alpha,\beta,\gamma)
  [y]_{s}[\alpha+s-y]_r=[\beta-y]_n[\gamma+n+y]_m~,
\end{eqnarray}
where $\alpha$, $\beta$ and $\gamma$ are arbitrary parameters. 
In terms of $A_{nm}^{rs}$ the ${\cal R}$-matrices for the 
above mentioned cases are given by
\begin{eqnarray}
  \label{rn0r0}
  {\cal R}_{n0}^{r0}{}_{;n'0}^{;r'0}({\scriptstyle {a\atop 
      \kappa}}|{\scriptstyle {a\atop \nu}};\varpi\bigr)
  &=&(-1)^{n-r}q^{-\frac{2}{3}\kappa\nu+(n-r)(\varpi^a+\kappa
    +n+r)-r'\kappa-r\nu-2rr'}
  \nonumber\\
  &&\hskip 2cm\times 
  A_{nn'}^{rr'}(\varpi^a+\nu+1,\varpi^a+1,\kappa)~,\nonumber\\
  {\cal R}_{0m}^{0s}{}_{;0m'}^{;0s'}({\scriptstyle {a\atop 
      \kappa}}|{\scriptstyle {a\atop \nu}};\varpi\bigr)
  &=&(-1)^{m-s}q^{-\frac{2}{3}\kappa\nu+(m-s)(\varpi^a+
    \varpi^{\bar a}+\kappa+m+s)-s'\kappa-s\nu-2ss'}
  \nonumber\\
  &&\hskip 2cm\times A_{mm'}^{ss'}(\varpi^a+\varpi^{\bar a}+\nu+1,
  \varpi^a+\varpi^{\bar a}+1,\kappa)~.
\end{eqnarray}

The solution to (\ref{anmrs}) has been obtained in ref. \cite{gs93} 
and is given by
\begin{eqnarray}
  \label{anmrssol}
  A_{nm}^{rs}(\alpha,\beta,\gamma)&\equiv&\frac{1}{[s]!}
  \frac{[\alpha]_{2s}}{[\alpha+s-1]_s}\frac{[\beta]_n
    [\gamma+n]_m}{[\alpha]_{n+m}}\nonumber\\
  &&\times\sum_{k=0}^s(-1)^k{s\choose l}_q
  \frac{[\alpha+s-1]_k}{[\alpha+n+m]_k}
  \frac{[\beta+n]_k}{[\beta]_k}
  \frac{[\gamma+n-k]_k}{[\gamma+n+m-k]_k} \nonumber\\
  &=&\frac{1}{[s]!}
  \frac{[\alpha]_{2s}}{[\alpha+s-1]_s}\frac{[\beta]_n
    [\gamma+n]_m}{[\alpha]_{n+m}}\nonumber\\
  &&\times{}_4F_3\left[\left.{\alpha+s-1,\atop \alpha+n+m,}
    {\beta+n,\atop \beta,}{1-\gamma-n,\atop 1-\gamma-n-m}
    {-s\atop {}}\:\right|\:q\:;\:1\:\right] ~,
\end{eqnarray}
where $\displaystyle{{n\choose m}_q\equiv\frac{[n]!}{[m]![n-m]!}}$ is the 
$q$-binomial coefficient and ${}_4F_3$ is a $q$-deformed 
hypergeometric function defined by
\begin{eqnarray}
  \label{qdhgf}
  {}_4F_3\left[\left.{a,\atop e,}{b,\atop f,}{c,\atop g}
    {d\atop {}}\right|\:q\:;z\right]
  \equiv\sum_{n=0}^\infty\frac{[a]_n[b]_n[c]_n[d]_n}{[e]_n[f]_n[g]_n[n]!}z^n
\end{eqnarray}
In Appendix C we give an elementary account of (\ref{anmrssol}). 
By making the substitution $y\rightarrow \beta-y$ in (\ref{anmrs}) and 
then defining new parameters by 
\begin{eqnarray}
  \label{adbdcd}
  \alpha'=\beta+\gamma~, \quad 
  \beta'=\beta~, \quad
  \gamma'=\alpha-\beta~,
\end{eqnarray}
we find that (\ref{anmrs}) can be inverted by the relation 
\begin{eqnarray}
  \label{aad}
  \sum_{n+n'=r+r'}A_{nn'}^{rr'}(\alpha,\beta,\gamma)
  A_{s's}^{n'n}(\alpha',\beta',\gamma')
  =\delta_{rs}\delta_{r's'}~.
\end{eqnarray}
We now show that $A_{nm}^{rs}$ satisfies
for some suitable choice of $c_{nm}$
\begin{eqnarray}
  \label{caca}
  c_{nn'}(\alpha,\beta,\gamma)A_{nn'}^{rr'}(\alpha,\beta,\gamma)
  =c_{r'r}(\alpha',\beta',\gamma')
  A_{r'r}^{n'n}(\alpha',\beta',\gamma')~.
\end{eqnarray}
To find $c_{nm}$, we need a property of the balanced 
$q$-hypergeometric function \cite{gs93}
\begin{eqnarray}
  \label{bqdhgf}
  &&{}_4F_3\left[\left.{a,\atop e,}{b,\atop f,}
      {c,\atop 1+a+b+c-e-f-n}
      {-n\atop {}}\right|\:q\:;1\right]\nonumber\\
  &&\hskip .5cm=\frac{[f-c]_n[e+f-a-b]_n}{[f]_n[e+f-a-b-c]_n}
  {}_4F_3\left[\left.{e-a,\atop e,}{e-b,\atop e+f-a-b,}
      {c,\atop 1+c-f-n}
      {-n\atop {}}\right|\:q\:;1\right]~, 
  \nonumber\\
\end{eqnarray}
where $n$ is a nonnegative integer. Using this together 
with the fact that ${}_4F_3$ defined by (\ref{qdhgf}) 
is symmetric in $a,b,c,d$ and in $e,f,g$, we obtain
\begin{eqnarray}
  \label{doubletr}
  &&{}_4F_3\left[\left.{\alpha+s-1,\atop \alpha+n+m,}
    {\beta+n,\atop \beta,}{1-\gamma-n,\atop 1-\gamma-n-m}
    {-s\atop {}}\:\right|\:q\:;\:1\:\right] \nonumber\\
  &&\hskip 1.5cm =\frac{[m]![\alpha-\beta+m]_n
    [\alpha+\gamma+2n+m-1]_m
    [\beta+\gamma+n+m]_n}{[r]![\alpha'-\beta'+r]_s
    [\alpha'+\gamma'+2s+r-1]_r[\beta'+\gamma'+r+s]_s}
  \nonumber\\&&\hskip 2.5cm\times 
  {}_4F_3\left[\left.{\alpha'+n-1,\atop \alpha'+r+s,}
    {\beta'+s,\atop \beta',}{1-\gamma'-s,\atop -\gamma'-r-s}
    {-n\atop {}}\:\right|\:q\:;\:1\:\right] ~.
\end{eqnarray}
We thus find $c_{nm}$ from (\ref{anmrssol}), (\ref{caca}) and 
(\ref{doubletr}) as 
\begin{eqnarray}
  \label{cnm}
  c_{nm}(\alpha,\beta,\gamma)
  =\frac{[\gamma]_n[\alpha-\beta]_m}{[n]![m]![\beta]_n
    [\alpha+\gamma+2n+m-1]_m[\beta+\gamma+n-1]_n
    [\beta+\gamma+2n]_m} ~.
\end{eqnarray}
where we have chosen the indeterminate multiplicative 
factor arbitrarily. 

In the above argument the explicit form of $A_{nm}^{rs}$ 
is used to show (\ref{caca}). It is possible to find 
$c_{nm}$ directly from (\ref{anmrs}) without referring 
to the details of $A_{nm}^{rs}$. This approach is 
promising since it can be applied for the cases where 
no explicit forms of the ${\cal R}$-matrix is not 
available. In the present case $c_{nm}$ can be defined as 
a solution to the following equation
\begin{eqnarray}
  \label{cseq}
  &&\sum_{n,n'} c_{nn'}(\alpha',\beta',\gamma')
  [y]_{n}[\alpha+n-y]_{n'}
  [\beta'-y']_{n}[\gamma'+n+y']_{n'} \nonumber\\
  &&\hskip 1cm =\sum_{n,n'} c_{nn'}(\alpha,\beta,\gamma)
  [y']_n[\alpha'+n-y']_{n'}
  [\beta-y]_n[\gamma+n+y]_{n'} ~,
\end{eqnarray}
where the sum should be taken over $n$ and $n'$ with $n+n'$ 
being fixed. 
A remarkable fact is that this becomes an identity 
with respect to $y$ and $y'$ for a suitable choice of the 
coefficient functions and $\alpha'$, $\beta'$, $\gamma'$ 
satisfying (\ref{adbdcd}). That (\ref{cseq}) leads to 
(\ref{caca}) can be shown by noting (\ref{anmrs}). 
As one may easily realize, eq. (\ref{cseq}) is 
nothing but (\ref{flocL2=0}) after a suitable identification 
of the parameters. Consequently, $c_{nm}$ can be expressed 
in terms of $C^a_{n0}$ given by (\ref{can0}). In fact it can be 
shown that the resultant expression obtained in this way 
coincides with (\ref{cnm}) up to a 
multiplicative factor. This immediately leads to the conclusion that 
the ${\cal R}$-matrices (\ref{rn0r0}) satisfy the locality 
constraints (\ref{locR2}). 

We next consider the exchange algebra (\ref{excalg}) for 
$b={\bar a}$. This is the second case where the ${\cal R}$-matrix 
is obtained systematically by the technique of truncated screening 
charges. It is straightforward to show that the ${\cal R}$-matrix 
must satisfy the relation
\begin{eqnarray}
  \label{Raab}
  &&\sum_{r+s+s'=n+m+m'\atop r'+s+s'=n'+m+m'}
  (-1)^rq^{-\frac{1}{3}\kappa\nu
    -s'(\varpi^a-r')+s(\varpi^{\bar a}+\nu+2r'+s'-r)
    -(r'+s')(r+s+s')}
  {\cal R}_{nm}^{rs}{}_{;n'm'}^{;r's'}
  ({\scriptstyle {a\atop 
      \kappa}}|{\scriptstyle {\bar a\atop \nu}};\varpi\bigr)\nonumber\\
  &&\hskip 1.5cm\times 
  [\varpi^a+1-r'-y]_r[\varpi^a+\varpi^{\bar a}+\nu+1+r'+s'-y]_s[y]_{s'}
  \nonumber\\
  &&\hskip 1.5cm\times 
  [\varpi^{\bar a}+1-z]_{r'}
  [\nu+r'+s'+z]_s
  [\varpi^a+\varpi^{\bar a}+1-z]_{s'} \nonumber\\
  &&=(-1)^nq^{-m'(\varpi^a+\kappa+2n+m-n')+m(\varpi^{\bar a}-n)+nn'-n'm'-mm'}
  \nonumber\\
  &&\hskip 1.5cm\times [\varpi^a+1-y]_n[\varpi^a+\varpi^{\bar a}+1-y]_m
  [\kappa+n+m+y]_{m'}\nonumber\\
  &&\hskip 1.5cm\times 
  [z]_m[\varpi^{\bar a}+1-n-z]_{n'}
  [\varpi^a+\varpi^{\bar a}+\kappa+1+n+m-z]_{m'} ~,
\end{eqnarray}
where $y$ and $z$ are arbitrary variables. This not only allows us to 
determine the ${\cal R}$-matrix recursively but also guarantees the 
full locality conditions (\ref{locR2}) for $b=\bar a$ when combined 
with (\ref{2ndlocc}) by the same reasoning explained above. The Toda 
exponential operators ${\rm e}^{\eta\kappa\lambda^a\cdot\varphi}$ 
with the expansion coefficients (\ref{canm}) which are only determined 
by using the truncated screening charges are indeed local with respect to 
${\rm e}^{\eta\nu\lambda^{\bar a}\cdot\varphi}$. 

\section{Discussion}
\label{sec:discussion}
\setcounter{equation}{0}

We have investigated $A_2$-Toda field theory in terms of the canonical 
free field. By introducing chiral schematic approach we have analyzed 
locality of the Toda exponentials. Locality turned out to impose 
nontrivial constraints on the elements of the $r$- and 
${\cal R}$-matrices. The main goal of sect. 2 was to establish that the 
classical solution (\ref{exacsol}) induces a canonical mapping from the 
interacting Toda system into a free theory. This has been achieved for 
Liouville and $A_2$-Toda theories. In ref. \cite{kn} the canonicity of the 
mapping was explicitly shown for Liouville theory within vector scheme. 
The efficiency of the chiral schematic description can be understood 
by noting that it can reproduce the same results in a systematic and 
simple way. Though the general treatment given in sect. 2 is not 
restricted to $A_2$-system, we have not been able to establish 
(\ref{cloc2}) for other Toda theories. Similar issue was studied by 
Babelon \cite{Babe} by using transfer matrix method and the chiral components 
of the canonical free fields were identified for general Toda theories. 
He worked in chiral scheme and the treatment of the boundary condition 
are somewhat different from ours. Periodic boundary condition was 
taken into account in ref. \cite{btb} and the quadratic Poisson algebra 
(\ref{qpoisalg}) was argued  for general Toda systems. 
As we have done, including the periodicity condition from the beginning 
brings about complications such as the extra zero-mode momentum dependences 
in the screening charges. This makes the explicit computations of the 
$r$-matrix rather cumbersome for higher rank theories. 

The presentation of the quantum theory in sect. 3 is restricted to 
$A_2$-system from the beginning. It could also be extended to general 
Toda theories. Main difficulty in generalizing our results is that 
tractable forms of the screening charges are not available for general 
cases. The property that the screening charges and the free field vertex 
appearing in the Toda exponential operator associated with a 
fundamental weight are mutually commuting is very crucial in our analysis 
and is considered to be universal for Toda theories. This may be 
established for the Toda exponential associated with the fundamental 
weight corresponding to the defining representation of $A_N$ algebra
since the set of the screening charges appearing there are a 
straightforward generalization of those of $A_1$- and $A_2$-systems. 
They are given by $A_{12\cdots k}$ ($k=1,\cdots,N$) for the left-moving 
sector with obvious notation. We also encounter new types of screening 
charges other than these. For instance in $A_3$-system the screening 
charges associated with $\lambda^2$, the six dimensional irreducible 
representation, are at the classical level
\begin{eqnarray}
  \label{a3sch}
  A_2~, \quad A_{21}~, \quad A_{23}~, \quad A_{213}+A_{231}~, 
  \quad A_{2132}+A_{2312}~.
\end{eqnarray}
The mutual commutatibity of their quantum counterparts is not obvious. 
Besides these points, the locality analysis in sect. 3 is not restricted 
to $A_2$-system since it relies essentially only on the fact that the 
Toda exponentials can be expanded as a bilinear form of chiral fields 
satisfying the quantum exchange algebra as given by (\ref{excalg}). 

Arbitrary exponential operators for $A_2$-system have been obtained in 
sect. 4 by generalizing the algebraic method of ref. \cite{fit96}. The locality 
requirement turned out to be so strong that only the truncated screening 
charges (\ref{tranys}) allowing two-dimensional quantum mechanical 
realization suffice to determine all the expansion coefficients of 
the exponential operators. We may stress that the algebraic method based 
on the quantum mechanical realization of the trancated screening charges 
is powerful in solving the locality conditions. This can be done without 
referring to the explicit forms of the quantum exchange algebra 
(\ref{excalg}). As has been argued in sect. 6, this method, however, fails 
in finding some elements of the ${\cal R}$-matrix. This simply reflects 
the fact that the truncation of the screening charges oversimplifies the 
full operator algebra. 

In conclusion it is possible to find exact operator solution for $A_2$ 
Toda field theory as a quantum deformation of the classical solution. 
The interacting to the free field canonical mapping is also realized  
at the quantum level.  Our algebraic method serves as an efficient tool 
in solving the locality conditions. We believe it to be applicable 
for the similar analysis of general Toda theories. 

\newpage
\appendix
\section{Poisson algebra among the chiral fields}
\label{sec:appA}
\setcounter{equation}{0}
In this appendix we collect the Poisson algebras among the chiral fields
$V^{a+}_\kappa\equiv{\rm e}^{\kappa\lambda^a\cdot\psi_+}$, $A_a$ and 
$A_{a\bar a}$. In order to simplify the formulae we put 
$\displaystyle{-\frac{\gamma^2}{4}}$ to unity in this appendix. One can 
recover the expressions for general values of $\gamma$ by the replacement 
$\displaystyle{\{~,~\}\rightarrow\Biggl(-\frac{\gamma^2}{4}\Biggr)^{-1}
\{~,~\}}$.
\begin{eqnarray}
  \label{aa1}
  \{V^{a+}_\kappa(x),V^{b+}_\nu(x')\}
  &=&\kappa\nu\lambda^a\cdot\lambda^b\epsilon(x-x')
  V^{a+}_\kappa(x)V^{b+}_\nu(x') ~,\nonumber\\
  \{V^{a+}_\kappa(x),A_a(x')\}&=&\kappa(\epsilon(x-x')A_a(x')+2C_{\alpha^a}
  {\cal E}_{\alpha^a}(x'-x)A_a(x))V^{a+}_\kappa(x) ~,\nonumber\\
  \{V^{a+}_\kappa(x),A_{\bar a}(x')\}&=&0 ~,\nonumber\\
  \{V^{a+}_\kappa(x),A_{a\bar a}(x')\}&=&\kappa(\epsilon(x-x')
  A_{a\bar a}(x')+2C_{\alpha^a+\alpha^{\bar a}}
  {\cal E}_{\alpha^a+\alpha^{\bar a}}(x'-x)A_{a\bar a}(x))V^{a+}_\kappa(x) ~,
  \nonumber\\
  \{V^{a+}_\kappa(x),A_{\bar aa}(x')\}&=&\kappa(\epsilon(x-x')A_{\bar aa}(x')
  +2C_{\alpha^a}{\cal E}_{\alpha^a}(x'-x)A_a(x)A_{\bar a}(x')
  \nonumber\\
  &&-2C_{\alpha^a+\alpha^{\bar a}}
  {\cal E}_{\alpha^a+\alpha^{\bar a}}(x'-x)A_{a\bar a}(x))V^{a+}_\kappa(x) ~,
  \nonumber\\
  \{A_a(x),A_a(x')\}&=&2\epsilon(x-x')A_a(x)A_a(x')\nonumber\\
  &&+2C_{\alpha^a}({\cal E}_{\alpha^a}(x'-x)(A_a(x))^2
  -{\cal E}_{\alpha^a}(x-x')(A_a(x'))^2) ~,\nonumber\\
  \{A_a(x),A_{a\bar a}(x')\}&=&\epsilon(x-x')A_a(x)A_{a\bar a}(x')
  +2C_{\alpha^a+\alpha^{\bar a}}
  {\cal E}_{\alpha^a+\alpha^{\bar a}}(x'-x)A_a(x)A_{a\bar a}(x) 
  \nonumber\\
  &&-2C_{\alpha^{\bar a}}{\cal E}_{\alpha^{\bar a}}(x'-x)A_{a\bar a}(x)
  A_a(x')-2C_{\alpha^a}{\cal E}_{\alpha^a}(x-x')A_a(x')A_{a\bar a}(x') ~,
  \nonumber\\ 
  \{A_{a\bar a}(x),A_{a\bar a}(x')\}&=&
  2\epsilon(x-x')A_{a\bar a}(x)A_{a\bar a}(x')\nonumber\\
  &&+2C_{\alpha^a+\alpha^{\bar a}}({\cal E}_{\alpha^a+\alpha^{\bar a}}(x'-x)
  (A_{a\bar a}(x))^2-{\cal E}_{\alpha^a+\alpha^{\bar a}}(x-x')
  (A_{a\bar a}(x'))^2) ~,\nonumber\\
  \{A_a(x),A_{\bar a}(x')\}&=&-\epsilon(x-x')A_a(x)A_{\bar a}(x')\nonumber\\
  &&-2C_{\alpha^{\bar a}}{\cal E}_{\alpha^{\bar a}}(x'-x)A_{a\bar a}(x)
  +2C_{\alpha^a}{\cal E}_{\alpha^a}(x-x')A_{\bar aa}(x') ~,
  \nonumber\\
  \{A_a(x),A_{\bar aa}(x')\}&=&\epsilon(x-x')A_a(x)A_{\bar aa}(x')
  +2C_{\alpha^a}{\cal E}_{\alpha^a}(x'-x)(A_a(x))^2
  A_{\bar a}(x') 
  \nonumber\\
  &&-2C_{\alpha^a+\alpha^{\bar a}}{\cal E}_{\alpha^a+\alpha^{\bar a}}(x'-x)
  A_a(x)A_{a\bar a}(x)  ~,\nonumber\\
  \{A_{a\bar a}(x),A_{\bar aa}(x')\}&=&2\epsilon(x-x')
  A_{a\bar a}(x)A_{\bar aa}(x') \nonumber\\
  &&+2(C_{\alpha^a}{\cal E}_{\alpha^a}(x'-x)A_{a\bar a}(x)
  -C_{\alpha^{\bar a}}{\cal E}_{\alpha^{\bar a}}(x-x')
  A_{\bar aa}(x'))A_a(x)
  A_{\bar a}(x')\nonumber\\
  &&-2C_{\alpha^a+\alpha^{\bar a}}({\cal E}_{\alpha^a+\alpha^{\bar a}}(x'-x)
  (A_{a\bar a}(x))^2-{\cal E}_{\alpha^a+\alpha^{\bar a}}(x-x')
  (A_{\bar aa}(x'))^2)~.\nonumber\\
\end{eqnarray}
Note that these are manifestly invariant under the shift 
$x\rightarrow x+2\pi$. All other brackets can be obtained by utilizing 
the symmetry $a\leftrightarrow\bar a$. 

The quadratic Poisson algebra satisfied by the chiral fields (\ref{a2cf}) 
can be easily found from the above Poisson brackets as 
\begin{eqnarray}
  \label{a2qpalg}
  \{\psi^+_0(x),\psi^+_0(x')\}&=&\frac{2}{3}\epsilon(x-x')
  \psi^+_0(x)\psi^+_0(x') ~,\nonumber\\
  \{\psi^+_0(x),\psi^+_1(x')\}&=&-\frac{1}{3}\epsilon(x-x')
  \psi^+_0(x)\psi^+_1(x') 
  -2C_{\alpha^a}{\cal E}_{\alpha^a}(x'-x)\psi^+_1(x)\psi^+_0(x')~,\nonumber\\
  \{\psi^+_0(x),\psi^+_2(x')\}&=&-\frac{1}{3}\epsilon(x-x')
  \psi^+_0(x)\psi^+_2(x')
  -2C_{\alpha^a+\alpha^{\bar a}}{\cal E}_{\alpha^a+\alpha^{\bar a}}(x'-x)
  \psi^+_2(x)\psi^+_0(x') ~,\nonumber\\
  \{\psi^+_1(x),\psi^+_1(x')\}&=&\frac{2}{3}\epsilon(x-x')
  \psi^+_1(x)\psi^+_1(x') ~,\nonumber\\
  \{\psi^+_1(x),\psi^+_2(x')\}&=&-\frac{1}{3}\epsilon(x-x')
  \psi^+_1(x)\psi^+_2(x')-2C_{\alpha^{\bar a}}{\cal E}_{\alpha^{\bar a}}(x'-x)
  \psi^+_2(x)\psi^+_1(x')~,\nonumber\\
  \{\psi^+_2(x),\psi^+_2(x')\}&=&\frac{2}{3}\epsilon(x-x')
  \psi^+_2(x)\psi^+_2(x') ~,\nonumber\\
  \{\psi^+_0(x),\psi^+_{\bar0}(x')\}&=&\frac{1}{3}\epsilon(x-x')
  \psi^+_0(x)\psi^+_{\bar0}(x') ~,\nonumber\\
  \{\psi^+_0(x),\psi^+_{\bar1}(x')\}&=&\frac{1}{3}\epsilon(x-x')
  \psi^+_0(x)\psi^+_{\bar1}(x') ~,\nonumber\\
  \{\psi^+_0(x),\psi^+_{\bar2}(x')\}&=&-\frac{2}{3}\epsilon(x-x')
  \psi^+_0(x)\psi^+_{\bar2}(x')
  -2C_{\alpha^a}{\cal E}_{\alpha^a}(x'-x)\psi^+_0(x)\psi^+_{\bar1}(x')
  \nonumber\\
  &&+2C_{\alpha^a+\alpha^{\bar a}}{\cal E}_{\alpha^a+\alpha^{\bar a}}(x'-x)
  \psi^+_2(x)\psi^+_{\bar0}(x') ~,\nonumber\\
  \{\psi^+_1(x),\psi^+_{\bar1}(x')\}&=&-\frac{2}{3}\epsilon(x-x')
  \psi^+_1(x)\psi^+_{\bar1}(x')
  +2C_{\alpha^a}{\cal E}_{\alpha^a}(x-x')\psi^+_0(x)\psi^+_{\bar2}(x')
  \nonumber\\
  &&-2C_{\alpha^{\bar a}}{\cal E}_{\alpha^{\bar a}}(x'-x)
  \psi^+_2(x)\psi^+_{\bar0}(x')~,
  \nonumber\\
  \{\psi^+_1(x),\psi^+_{\bar2}(x')\}&=&\frac{1}{3}\epsilon(x-x')
  \psi^+_1(x)\psi^+_{\bar2}(x')~,\nonumber\\
  \{\psi^+_2(x),\psi^+_{\bar2}(x')\}&=&\frac{1}{3}\epsilon(x-x')
  \psi^+_2(x)\psi^+_{\bar2}(x') ~.
\end{eqnarray}
One can read off the $r$-matrix (\ref{a2rmat}) from these 
Poisson brackets.

\section{Basic Operator Algebra in $A_2$-Toda Theory}
\label{sec:appB}
\setcounter{equation}{0}

This appendix concerns the basic exchange algebra satisfied by 
the chiral vertex operators $V^{a+}_\kappa$ and the screening 
charges $A_a$, $A_{a\bar a}$. 
\begin{eqnarray}
  \label{oep}
  V^{a+}_\kappa(x)V^{a+}_\nu(x')
  &=&q^{-\frac{2}{3}\kappa\nu\epsilon(x-x')}V^{a+}_\kappa(x')
  V^{a+}_\kappa(x) ~,\nonumber\\
   V^{a+}_\kappa(x)A_a(x')
  &=&\Biggl\{q^{-\kappa\epsilon(x-x')}\frac{[\varpi^a+\kappa+1]}{[\varpi^a+1]}
  A_a(x')-q^{-(\varpi^a+\kappa+1)\epsilon(x-x')}\frac{[\kappa]}{[\varpi^a+1]}
  A_a(x)\Biggr\}V^{a+}_\kappa(x) ~,\nonumber\\
   A_a(x)V^{a+}_\kappa(x')
  &=&V^{a+}_\kappa(x')\Biggl\{q^{-\kappa\epsilon(x-x')}
  \frac{[\varpi^a-\kappa+1]}{[\varpi^a+1]}
  A_a(x)+q^{(\varpi^a-\kappa+1)\epsilon(x-x')}\frac{[\kappa]}{[\varpi^a+1]}
  A_a(x')\Biggr\} ~,\nonumber\\
   V^{a+}_\kappa(x)A_{a\bar a}(x')
  &=&\Biggl\{q^{-\kappa\epsilon(x-x')}
  \frac{[\varpi^a+\varpi^{\bar a}+\kappa+1]}{[\varpi^a+\varpi^{\bar a}+1]}
  A_{a\bar a}(x') \nonumber\\
  && -q^{-(\varpi^a+\varpi^{\bar a}+\kappa+1)\epsilon(x-x')}
  \frac{[\kappa]}{[\varpi^a+\varpi^{\bar a}+1]}
  A_{a\bar a}(x)\Biggr\}V^{a+}_\kappa(x) ~,\nonumber\\
  A_{a\bar a}(x)V^{a+}_\kappa(x')
  &=&V^{a+}_\kappa(x')\Biggl\{q^{-\kappa\epsilon(x-x')}
  \frac{[\varpi^a+\varpi^{\bar a}-\kappa+1]}{[\varpi^a+\varpi^{\bar a}+1]}
  A_{a\bar a}(x) \nonumber\\
  && +q^{(\varpi^a+\varpi^{\bar a}-\kappa+1)\epsilon(x-x')}
  \frac{[\kappa]}{[\varpi^a+\varpi^{\bar a}+1]}
  A_{a\bar a}(x')\Biggr\}\nonumber\\
  A_a(x)A_a(x')
  &=&q^{-2\epsilon(x-x')}A_a(x')A_a(x)
  -\frac{q^{-(\varpi^a+3)\epsilon(x-x')}}{[\varpi^a+2]}(A_a(x))^2
  +\frac{q^{(\varpi^a+1)\epsilon(x-x')}}{[\varpi^a+2]}(A_a(x'))^2 ~,
  \nonumber\\
  A_a(x)A_{a\bar a}(x')
  &=&q^{-\epsilon(x-x')}
  \frac{[\varpi^a+1][\varpi^{\bar a}-1][\varpi^a+\varpi^{\bar a}+2]}
  {[\varpi^a+2][\varpi^{\bar a}][\varpi^a+\varpi^{\bar a}+1]}
  A_{a\bar a}(x')A_a(x)\nonumber\\
  && +\frac{q^{-(\varpi^{\bar a}+1)\epsilon(x-x')}}{[\varpi^{\bar a}]}
  A_a(x')A_{a\bar a}(x)
  -\frac{q^{-(\varpi^a+\varpi^{\bar a}+2)\epsilon(x-x')}}
  {[\varpi^a+\varpi^{\bar a}+1]}A_a(x)A_{a\bar a}(x)\nonumber\\
  &&+\frac{q^{(\varpi^a+1)\epsilon(x-x')}}{[\varpi^a+2]}
  A_a(x')A_{a\bar a}(x')  ~,\nonumber\\
  A_{a\bar a}(x)A_a(x')
  &=&q^{-\epsilon(x-x')}
  \frac{[\varpi^a+2][\varpi^{\bar a}+1][\varpi^a+\varpi^{\bar a}+1]}
  {[\varpi^a+1][\varpi^{\bar a}][\varpi^a+\varpi^{\bar a}+2]}
  A_a(x')A_{a\bar a}(x)\nonumber\\
  && -\frac{q^{(\varpi^{\bar a}-1)\epsilon(x-x')}}{[\varpi^{\bar a}]}
  A_{a\bar a}(x')A_a(x)
  -\frac{q^{-(\varpi^a+2)\epsilon(x-x')}}
  {[\varpi^a+1]}A_a(x)A_{a\bar a}(x)\nonumber\\
  &&+\frac{q^{(\varpi^a+\varpi^{\bar a}+1)\epsilon(x-x')}}
  {[\varpi^a+\varpi^{\bar a}+2]}
  A_a(x')A_{a\bar a}(x')  ~,\nonumber\\
  A_{a\bar a}(x)A_{a\bar a}(x')
  &=&q^{-2\epsilon(x-x')}A_{a\bar a}(x')A_{a\bar a}(x)
  -\frac{q^{-(\varpi^a+\varpi^{\bar a}+3)\epsilon(x-x')}}
  {[\varpi^a+\varpi^{\bar a}+2]}(A_{a\bar a}(x))^2 \nonumber\\
  && +\frac{q^{(\varpi^a+\varpi^{\bar a}+1)\epsilon(x-x')}}
  {[\varpi^a+\varpi^{\bar a}+2]}(A_{a\bar a}(x'))^2  ~,\nonumber\\
  V^{a+}_\kappa(x)V^{\bar a+}_\kappa(x')
  &=&q^{-\frac{1}{3}\kappa\nu\epsilon(x-x')}V^{\bar a+}_\kappa(x')
  V^{a+}_\kappa(x) ~,\nonumber\\
  V^{a+}_\kappa(x)A_{\bar a}(x')&=&A_{\bar a}(x')V^{a+}_\kappa(x) ~,\nonumber\\
  V^{a+}_\kappa(x)A_{\bar a a}(x')
  &=&\Biggl\{q^{-\kappa\epsilon(x-x')}\frac{[\varpi^a+\kappa]
    [\varpi^a+\varpi^{\bar a}+\kappa+1]}{[\varpi^a+\varpi^{\bar a}+1]
    [\varpi^{\bar a}]}A_{\bar aa}(x') \nonumber\\
  &&-q^{-(\varpi^a+\kappa)\epsilon(x-x')}
  \frac{[\kappa][\varpi^a+\varpi^{\bar a}+1]}{[\varpi^a][\varpi^a+1]}
  A_{\bar a}(x')A_a(x) \nonumber\\
  &&+q^{-(\varpi^a+\varpi^{\bar a}+\kappa+1)\epsilon(x-x')}
  \frac{[\kappa][\varpi^a+\kappa]}{[\varpi^a+\varpi^{\bar a}+1]
    [\varpi^{\bar a}+1]}A_{a\bar a}(x)\Biggr\}V^{a+}_\kappa(x)  ~,\nonumber\\
  A_{a\bar a}(x)V^{\bar a+}_\kappa(x')
  &=&V^{\bar a+}_\kappa(x')\Biggl\{q^{-\kappa\epsilon(x-x')}
  \frac{[\varpi^a+\varpi^{\bar a}-\kappa+1]
    [\varpi^{\bar a}-\kappa+1]}{[\varpi^a+\varpi^{\bar a}+1]
    [\varpi^{\bar a}+1]}
  A_{a\bar a}(x)\nonumber\\
  &&+q^{(\varpi^{\bar a}-\kappa+1)\epsilon(x-x')}
  \frac{[\kappa][\varpi^a+\varpi^{\bar a}-\kappa+1]}{[\varpi^a]
    [\varpi^{\bar a}+1]}A_{\bar a}(x')A_a(x)
  \nonumber\\
  &&-q^{(\varpi^a+\varpi^{\bar a}-\kappa+1)\epsilon(x-x')}
  \frac{[\kappa][\varpi^{\bar a}+1]}{[\varpi^a]
    [\varpi^a+\varpi^{\bar a}+1]}A_{a\bar a}(x)\Biggr\}  ~,\nonumber\\
  A_a(x)A_{\bar a}(x')
  &=&q^{\epsilon(x-x')}\frac{[\varpi^a+1]
    [\varpi^{\bar a}]}{[\varpi^a][\varpi^{\bar a}+1]}A_{\bar a}(x')
  A_a(x)\nonumber\\
  &&+q^{-\varpi^{\bar a} \epsilon(x-x')}
  \frac{[\varpi^a+1]}{[\varpi^a+\varpi^{\bar a}+1][\varpi^{\bar a}+1]}
  A_{a\bar a}(x)\nonumber\\
  &&-q^{(\varpi^a+1)\epsilon(x-x')}
  \frac{[\varpi^{\bar a}]}{[\varpi^a][\varpi^a+\varpi^{\bar a}+1]}
  A_{\bar aa}(x')  ~,\nonumber\\
  A_a(x)A_{\bar aa}(x')
  &=&q^{-\epsilon(x-x')}\frac{[\varpi^a+1]
    [\varpi^a+\varpi^{\bar a}+2]}{[\varpi^a]
    [\varpi^a+\varpi^{\bar a}+1]}
  A_{\bar aa}(x')A_a(x)\nonumber\\
  &&+q^{-(\varpi^a+\varpi^{\bar a}+2)\epsilon(x-x')}
  \frac{[\varpi^a+1]}{[\varpi^a+\varpi^{\bar a}+1][\varpi^{\bar a}+1]}
  A_a(x)A_{a\bar a}(x)\nonumber\\
  &&-q^{-(\varpi^a+1)\epsilon(x-x')}
  \frac{[\varpi^a+\varpi^{\bar a}+2]}{[\varpi^a][\varpi^{\bar a}+1]}
  A_{\bar a}(x')(A_a(x))^2 ~,\nonumber\\
  A_{a\bar a}(x)A_{\bar a}(x')
  &=&q^{-\epsilon(x-x')} \frac{[\varpi^{\bar a}+2]
    [\varpi^a+\varpi^{\bar a}+1]}{[\varpi^a+\varpi^{\bar a}+2]
    [\varpi^{\bar a}+3]}
  A_{\bar a}(x')A_{a\bar a}(x)\nonumber\\
  &&-q^{(\varpi^a+\varpi^{\bar a}+1)\epsilon(x-x')}
  \frac{[\varpi^{\bar a}+2]}{[\varpi^a-1][\varpi^a+\varpi^{\bar a}+2]}
  A_{\bar a}(x')A_{\bar aa}(x')\nonumber\\
  &&+q^{(\varpi^{\bar a}+2)\epsilon(x-x')}
  \frac{[\varpi^a+\varpi^{\bar a}+1]}{[\varpi^a-1][\varpi^{\bar a}+3]}
  (A_{\bar a}(x'))^2A_a(x)  ~,\nonumber\\
  A_{a\bar a}(x)A_{\bar aa}(x')
  &=&q^{-2\epsilon(x-x')}\frac{[\varpi^a+1]
    [\varpi^{\bar a}+1]}{[\varpi^a][\varpi^{\bar a}+2]}
  A_{\bar aa}(x')A_{a\bar a}(x)\nonumber\\
  &&+q^{-(\varpi^a+\varpi^{\bar a}+3)
    \epsilon(x-x')}\frac{[\varpi^a+1]}{[\varpi^a+\varpi^{\bar a}+2]
    [\varpi^{\bar a}+2]}(A_{a\bar a}(x))^2\nonumber\\
  &&-q^{(\varpi^a+\varpi^{\bar a}+1)\epsilon(x-x')}
  \frac{[\varpi^{\bar a}+1]}{[\varpi^a][\varpi^a+\varpi^{\bar a}+2]}
  (A_{\bar aa}(x'))^2\nonumber\\
  &&-q^{-(\varpi^a+2)\epsilon(x-x')}
  \frac{[\varpi^a+\varpi^{\bar a}+1]
    [\varpi^a+\varpi^{\bar a}+3]}{[\varpi^a]
    [\varpi^a+\varpi^{\bar a}+2][\varpi^{\bar a}+3]}
  A_{\bar a}(x')A_a(x)A_{a\bar a}(x)\nonumber\\
  &&+q^{\varpi^{\bar a} \epsilon(x-x')}
  \frac{[\varpi^a+\varpi^{\bar a}+1]
    [\varpi^a+\varpi^{\bar a}+3]}{[\varpi^a-1]
    [\varpi^a+\varpi^{\bar a}+2][\varpi^{\bar a}+2]}
  A_{\bar a}(x')A_{\bar aa}(x')A_a(x)\nonumber\\
  &&-q^{-(\varpi^a-\varpi^{\bar a}-1)\epsilon(x-x')} 
    \frac{[\varpi^a+\varpi^{\bar a}+1]
      [\varpi^a+\varpi^{\bar a}+3]}{[\varpi^a-1]
      [\varpi^a][\varpi^{\bar a}+2][\varpi^{\bar a}+3]}
    (A_{\bar a}(x'))^2(A_a(x))^2~.
\end{eqnarray}
All other formulae can be obtained from these by the 
interchange $a\leftrightarrow \bar a$ of the indices. 
It can be easily verified that in the classical limit 
defined in sect. \ref{sec:qteo} these reduce to the 
classical results (\ref{aa1}) if one recalls the 
definitions (\ref{a}), (\ref{cscop}) and identifies the 
Poisson brackets by
\begin{eqnarray}
  \label{qtoclass}
  \{F,G\}&=&\lim_{g\rightarrow 0}\;\frac{1}{2\pi ig}[F,G]~.
\end{eqnarray}

\section{Proof of $A_{nm}^{rs}(\alpha,\beta,\gamma)$}
\label{sec:proof}
\setcounter{equation}{0}

In this appendix we give an elementary proof of 
(\ref{anmrssol}) by induction with respect to $s$ 
for fixed $n$, $m$. Let us define 
\begin{eqnarray}
  \label{zl}
  &&A_s\equiv A_{nm}^{rs}(\alpha,\beta,\gamma)~, \qquad 
  z_l\equiv\frac{[\beta+n]_l}{[\beta]_l}
  \frac{[\gamma+n-l]_l}{[\gamma+n+m-l]_l} ~, \nonumber\\
  &&a^a_l\equiv(-1)^l{s\choose l}_q
  \frac{[\alpha+s-1]_l}{[\alpha+n+m]_l} ~, \qquad
  c_s\equiv\frac{1}{[s]!}
  \frac{[\alpha]_{2s}}{[\alpha+s-1]_s}
  \frac{[\beta]_n[\gamma+n]_m}{[\alpha]_{n+m}} ~,
\end{eqnarray}
then (\ref{anmrssol}) can be rewritten as 
\begin{eqnarray}
  \label{rwanmrs}
  A_s=c_s\sum_{l=0}^{s} 
  a^s_lz_l ~. \qquad (0\le s\le n+m)
\end{eqnarray}
This holds true for $s=0$ as can be easily seen by 
putting $y=0$ in (\ref{anmrs}). We assume that 
(\ref{rwanmrs}) is correct for $s\le k$ with some 
positive integer $k<n+m$. Then $A_{k+1}$ can be 
found from (\ref{anmrs}) for $y=-k-1$ as 
\begin{eqnarray}
  \label{akp1}
  A_{k+1}=\sum_{l=0}^{k}\Biggl\{\sum_{s=l}^{k}(-1)^{k-s}
  \frac{[k-s+2]_s[\alpha+k+s+1]_{n+m-s}}{%
    [k+1]![\alpha+2k+2]_{n+m-k-1}}c_sa^s_l\Biggr\}z_l
  +c_{k+1}a^{k+1}_{k+1}z_{k+1} ~.
\end{eqnarray}
Hence we must show 
\begin{eqnarray}
  \label{akp1-2}
  \sum_{s=l}^{k}(-1)^{k-s}
  \frac{[k-s+2]_s[\alpha+k+s+1]_{n+m-s}}{%
    [k+1]![\alpha+2k+2]_{n+m-k-1}}c_sa^s_l=c_{k+1}a^{k+1}_l~.
\end{eqnarray}
This can be stated equivalently as a formula
\begin{eqnarray}
  \label{formula}
  \sum_{s=0}^N(-1)^{N-s}\frac{[N-s+2]_s}{[s]!}
  [\alpha+2s][\alpha]_s[\alpha+N+s+2]_{N-s}=[\alpha]_{N+1}~.
\end{eqnarray}
It can be established as follows: If we define a function 
$h(\alpha)$ by the lhs of (\ref{formula}), then it vanishes 
for $\alpha=0,-1,\cdots,-N$. This immediately implies that 
$h(\alpha)$ is proportional to $[\alpha]_{N+1}$. 
The equality can be seen by putting $\alpha=-2N-1$. 

Eqs. (\ref{akp1}) and (\ref{akp1-2}) imply that (\ref{rwanmrs}) 
is correct for $s=k+1$. This completes the proof of 
(\ref{rwanmrs}). 

\eject
\end{document}